\documentclass[12pt,preprint]{aastex}

\usepackage{graphicx}
\usepackage{comment}
\usepackage{amsmath}
\usepackage{epstopdf}

\def\beq{\begin{equation}} 
\def\eeq{\end{equation}} 
\def\beqar{\begin{eqnarray}} 
\def\eeqar{\end{eqnarray}}

\def\fun#1#2{\lower3.6pt\vbox{\baselineskip0pt\lineskip.9pt
  \ialign{$\mathsurround=0pt#1\hfil##\hfil$\crcr#2\crcr\sim\crcr}}}

\begin{document} 

\title{Probing the Cosmic Gamma-Ray Burst Rate with Trigger Simulations of the \textit{Swift} Burst Alert Telescope}

\author{Amy Lien$^{1}$,Takanori Sakamoto$^2$,Neil Gehrels$^3$,David M. Palmer$^4$,\\Scott~D.~Barthelmy$^3$,Carlo Graziani$^{5,6}$,John~K.~Cannizzo$^{7,8}$}
\affil{$^1$NASA Postdoctoral Program Fellow, Goddard Space Flight Center, Greenbelt, MD 20771, USA}
\affil{$^2$Department of Physics and Mathematics, College of Science and Engineering, Aoyama Gakuin University, 5-10-1 Fuchinobe, Chuo-ku, Sagamihara-shi, Kanagawa 252-5258, Japan}  
\affil{$^3$NASA Goddard Space Flight Center, Greenbelt, MD 20771, USA}
\affil{$^4$Los Alamos National Laboratory, B244, Los Alamos, NM, 87545, USA}
\affil{$^5$Astronomy Department, The University of Chicago, Chicago, IL 60637}
\affil{$^6$Flash Center Computational Science, The University of Chicago, Chicago, IL 60637}
\affil{$^7$Center for Research and Exploration in Space Science and Technology (CRESST) and NASA Goddard Space Flight Center, Greenbelt, MD 20771, USA}
\affil{$^8$Department of Physics, University of Maryland, Baltimore County, Baltimore, MD 21250, USA}

\begin{abstract}

The gamma-ray burst (GRB) rate is essential for revealing the connection between GRBs, supernovae and stellar evolution. Additionally, the GRB rate at high redshift provides a strong probe of star formation history in the early universe. While hundreds of GRBs are observed by {\it Swift}, it remains difficult to determine the intrinsic GRB rate due to the complex trigger algorithm of {\it Swift}. Current studies of the GRB rate usually approximate the {\it Swift} trigger algorithm by a single detection threshold. However, unlike the previously flown GRB instruments, {\it Swift} has over 500 trigger criteria based on photon count rate and additional image threshold for localization. To investigate possible systematic biases and explore the intrinsic GRB properties, we develop a program that is capable of simulating all the rate trigger criteria and mimicking the image threshold. Our simulations show that adopting the complex trigger algorithm of {\it Swift} increases the detection rate of dim bursts. As a result, our simulations suggest bursts need to be dimmer than previously expected to avoid over-producing the number of detections and to match with {\it Swift} observations. Moreover, our results indicate that these dim bursts are more likely to be high redshift events than low-luminosity GRBs. This would imply an even higher cosmic GRB rate at large redshifts than previous expectations based on star-formation rate measurements, unless other factors, such as the luminosity evolution, are taken into account.  The GRB rate from our best result gives a total number of $4571^{+829}_{-1584}$ GRBs per year that are beamed toward us in the whole universe.

{\bf Special note (2015.05.16): This new version incorporates an erratum. All the GRB rate normalizations ($R_{\rm GRB}(z=0)$) should be a factor of 2 smaller than previously reported. Please refer to the Appendix for more details. All the values are corrected in this version. We sincerely apologize for the mistake, and for not noticing it earlier.}

{\bf Special note (2016.03.22): There was a typo in Eq.~\ref{eq:Yonetoku_mod}, which is fixed in this version. This was a typo in the latex file, and thus results and numbers are not affected.}

\end{abstract}

\noindent

\section{Introduction}

\label{sect:intro}

Gamma-ray bursts (GRBs) are one of the most energetic phenomena in the universe.
Observationally, GRBs are usually categorized as long and short bursts, 
with an empirical separation of two seconds in their observational durations. 
The burst duration is often referred as $T_{90}$, which is
the light curve period that encloses $90\%$ of the burst photons.
Long GRBs are expected to result from explosions of massive stars
with powerful central engines such as a black holes \citep[e.g.,][]{Heger03}.
Additionally, observations have shown that at least some long GRBs are connected to Type Ic supernovae
\citep[e.g.,][]{Galama98, Bloom02, DellaValle03, Stanek03,Thomsen04,Campana06, Starling11, Berger11, Vergani11, Sparre11, Melandri12}. 
Short GRBs are believed to originate from
compact-object mergers due to their different host galaxy properties and non-detections of the accompanied supernovae 
\citep[e.g.,][]{Eichler89, Nakar07,Fong10, Fong13}.
Here we will focus on the long GRBs, since they are closely related to 
massive stars, and hence trace the star-formation history more directly.
Throughout the paper, GRBs refer to long bursts unless otherwise specified.

Due to their extraordinary luminosities, 
GRBs provide a unique and independent method for measuring the cosmic star-formation rate (SFR), especially at large redshifts
where it becomes difficult for other methods.
Many efforts have been done to map out the intrinsic cosmic GRB rate as a function of redshift from 
current observations \citep[e.g.,][]{Guetta07_Ibc,Guetta07_sfr,Yuksel08,Kistler08_SFR,Butler10,Robertson12,Pelangeon08,Salvaterra09,Campisi10,Wanderman10,Virgili11, Qin11,Salvaterra12,Coward13,Kanaan13}.
Results from these studies suggest that the cosmic GRB rate generally follows the shape of the cosmic SFR
at low redshift. However, at large redshift (z $\gtrsim$ 5), several groups have suggested 
a higher GRB rate than previously thought based on measurements of the SFR.
\citep[e.g.,][]{Le07,  Yuksel08, Kistler09, Butler10, Ishida11,Tanvir12,Jakobsson12}.
For example, \citet{Kistler09} conclude that at high redshift the GRB rate does not trace 
the commonly adopted SFR from \citet{Hopkins06}.
These authors further state that the higher SFR implied by the GRB rate
can be explained when including the star formation from undetectable galaxies at the faint end of the UV luminosity function,
and therefore stars alone are sufficient to explain the reionization in the early universe \citep{Kistler13}.
\citet{Tanvir12} used the high-redshift GRBs detected by {\it Swift} for 
locating their host galaxies and performing observations via the Hubble Ultra-Deep Field.
Based on the non-detection of the host galaxies, they put an upper limit on the SFR of these galaxies 
and came to a similar conclusion as \citet{Kistler09}, that most of the star formation at high redshift comes from low-luminosity galaxies.

To estimate the intrinsic GRB rate from current observations, one needs to 
convert the observed rate back to the intrinsic rate based on GRB luminosity distribution,
the survey sensitivity, and other GRB characteristics such as the GRB spectral information,
burst shapes/durations, and the beaming angles.
Recently, several groups have done some careful examinations of the cosmic GRB rate via Monte Carlo approaches 
with adjustable parameters in the GRB rate and luminosity distribution for fitting \citep{Butler10, Wanderman10, Qin11, Virgili11}. 
Some studies also adopt additional correlation between the GRB characteristics 
(e.g., luminosities, the peak energy $E^{\rm src}_{\rm peak}$ of the burst spectrum)
in their Monte Carlo search \citep[e.g.,][]{Butler10}.

Most of these studies focus on searching for the intrinsic GRB rate that are beamed toward us,
because we can only observe GRBs that are pointed at us, and it is extremely difficult to 
measure the beaming angle. Despite all the effort, there remain large uncertainties in
the beaming angle and the beaming factor, which is the fraction of GRBs that are beamed toward us. 
The beaming factor
can range from $\sim 50$ to $\sim 500$ \citep[e.g.,][]{Frail01,Guetta05}.
To avoid involving further uncertainties in our calculation, 
we also focus on finding the intrinsic GRB rate for GRBs that are beamed at us. 
All the GRB rates in this paper refer to GRBs that pointed toward us,
unless specifically noted.
 
Most recent studies concerning GRB rate must
estimate a sensitivity for the
survey of each detector under consideration.
A single detection threshold is most commonly used for estimating the survey sensitivity,
in which a flux (or photon count rate) limit is used as the instrument limit
and a GRB with flux (or photon count rate) above the limit is considered as a trigger event.
This is generally a good assumption for GRB instruments prior to {\it Swift}.
Unlike previously flown GRB instruments, however,
{\it Swift} adopts a much more complex trigger algorithm in order to
maximize the number of GRB detections \citep{Band06}.
The Burst Alert Telescope (BAT) of {\it Swift} adopts over 500 ``rate trigger'' criteria, based on photon count rates,
and additional `` image threshold''  based on the real image generated for further confirmation and localization
\citep{Barthelmy05, Fenimore03, Fenimore04, McLean04, Palmer04}.
Therefore, a single-detection threshold approximation might not be
proper for correctly estimating the survey sensitivity of {\it Swift}.

A few studies thus adopt more sophisticated approximations of the complex trigger algorithm of {\it Swift}.
For example, \citet{Butler10} used an empirical combination of photon counts and pulse durations to approximate
the signal-to-noise ratio cut of {\it Swift} detections.
\citet{Wanderman10} introduced an empirical probability function of detectability for 
bursts with flux close to the assumed single flux threshold. Additionally,
these authors also adopted a empirical function to account for 
the probability of redshift measurements.

Despite all the deliberate methods adopted to explore the intrinsic GRB rate,
it remains difficult to quantify the selection biases from the complex BAT trigger algorithm.
We hence proceed with an alternative approach by
actually simulating the BAT trigger algorithm, including all the rate trigger criteria and 
image threshold.
We will use this ``BAT-trigger simulator'' to search for cosmic GRB rate and luminosity
distributions that generate a mock-triggered sample that 
can reproduce the observational GRB characteristics. 

This paper is organized as follow: 
Section \ref{sect:trigger_algorithm} describes the complex trigger algorithm adopted by the BAT.
Section \ref{sect: BAT-simulator} explains the method we use in our 
simulations, including generating GRB light curves based on input burst properties and simulating the BAT trigger algorithm.
Section \ref{sect:BAT_observables} summarizes the observational GRB samples we adopted 
to be compared with our simulations. 
Section \ref{sect:rate} presents our best finding of the cosmic GRB rate and luminosity functions
that generate results which match the best with the observational GRB characteristics.
Section \ref{sect:Epeak} explores the consequences of adopting different distributions of GRB spectra ($E^{\rm src}_{\rm peak}$),
including spectral evolution.
Section \ref{sect:biases} discusses the advantages of using the complex BAT trigger algorithm, 
and the selection biases introduced by using a single trigger criterion.
Section \ref{sect:detection_rate} shows the GRB detection fraction as a function of redshift, 
estimated by the BAT-trigger simulator. 
Section \ref{sect:sfr} compares our best-fit cosmic GRB rate with the cosmic SFR.
Section \ref{sect:lum_evo} explores the possibility of luminosity evolution.
Finally, the results and their implications are presented in Sect.~\ref{sect:conclusions}.
Throughout this study, we adopt a standard flat $\Lambda$CDM 
model with $\Omega_{\rm m} = 0.274$, $\Omega_{\Lambda}=0.726$,
and $H_0 = 70.5 \, \rm km \, s^{-1} \, Mpc^{-1}$
\citep{wmap07}.

\section{\textit{Swift}-BAT's trigger algorithm}
\label{sect:trigger_algorithm}

During the observational process, BAT constantly calculates the signal-to-noise ratio of the detected light curves based on photon count rates in some 
assigned foreground and background time periods. Each foreground and background period are separated by an ``elapse time'' period to reduce the chance 
of the background photons being contaminated by the foreground photons. 
An event is ``triggered'' when the signal-to-noise ratio calculated
within a given interval exceeds the assigned signal-to-noise ratio
threshold.
We adopt Eq.~3 in \citet{Graziani03} for calculating the signal-to-noise ratio. This is generally the same equation used by the BAT flight software, except 
the equation in BAT's algorithm includes  an extra term to prevent errors occurring when there are zeros in the denominator, that is, zero photons in the background light curves \citep{Fenimore03}.
In our simulation we will not have the case with zero photons in the light curves, hence we do not need to include such term.

BAT has 674 different rate trigger criteria.
Each rate trigger criterion adopts a different 
signal-to-noise ratio threshold, different foreground, background, and elapse time periods for calculating the signal-to-noise ratio,
and covers a different energy band \citep{Fenimore03}. The complex trigger algorithm is implemented to increase the chance of correctly bracketing the GRB pulse shape
and hence successfully find a GRB. 
After an event is triggered by one of the rate-trigger criteria, an image will be generated for further confirmation and localization.
During this imaging process, an additional signal-to-noise ratio based on the real image will be calculated. The rate-triggered event will be confirmed 
as a real detection if (1) the image signal-to-noise ratio is higher than the image threshold (signal-to-noise ratio $\gtrsim 6.5$), and (2) the event is compared with current on-board sky catalog and no known
source is found to be at the event location.
Out of all the rate trigger criteria, 
there are 180 criteria that have extremely short foreground periods ($\leq 0.064$ sec).
These short trigger criteria are particularly aimed for detecting GRBs with very short durations.
Since we focus on long GRBs in this paper, 
we do not include these short trigger criteria in our simulations.

A throttling process is implemented to reduce the number of trigger criteria if 
the CPU is getting too far behind in processing the data.
Additionally, BAT will also temporarily suspend the trigger process 
during some particular occasions, such as satellite slewing,  when there are too many hot pixels that
significantly reduce the detection sensitivity,
or when the satellite is passing through the South Atlantic Anomaly region,
which is an area of the Earth's orbit that contains much higher background flux. 
In our simulation, we only consider the trigger process during normal situations. 
However, we take into account these ``deadtimes'' of observations 
by multiplying the simulated trigger rate by a fraction of the survey time ($\sim 90\%$)
when the telescope is actually performing the observations.

In addition to the rate trigger process, a burst could also be found by an independent ``image trigger'' process,
in which a regular image is generated by BAT every minute or longer to search for dim GRBs that are missed by 
the rate trigger. A typical image trigger process generates images
from light curve periods of one and five minutes.
However, when a rate trigger is active, the flight software also generates images 
from extended periods with increments of 8 seconds. 
That is, images from light curve periods of (64 + multiples of 8) seconds (i.e., 72s, 80s, 88s...etc) 
will also be created, until the extension time reaches another minute.

\section{Simulations of GRB light curves and BAT trigger algorithm}
\label{sect: BAT-simulator}

In order to search for a more robust cosmic GRB rate and to study possible systematic effects due to the complex trigger algorithm of BAT, 
we developed a code that is capable of creating mock observed GRB light curves based on adjustable burst properties, 
and simulating the BAT trigger algorithm for the first time,
including simulating hundreds of rate trigger criteria and mimicking the image threshold. This program contains three main parts, as described in the following subsections. 

\subsection {Creating the mock light curves in the GRB rest frame based on assumed GRB characteristics} 
\label{sect:lc_rest}
We create a sample of mock GRB light curves in the rest frame based on several input GRB properties, which include: 
\begin{enumerate}
\item Cosmic GRB rate. We adopt the functional form in \citet{Wanderman10}, which
assumes a simple broken power law shape that generally follows the shape of the cosmic SFR, with adjustable parameters. That is,
the cosmic GRB rate increases to some redshift $z_1$, and decreases afterward, as shown in Eq.~\ref{eq:GRBrate}.
\beq
\label{eq:GRBrate}
R_{\rm GRB}(z) = R_{\rm GRB}(z=0) \ \left \{  \begin{array}{ll}
(1+z)^{n_1}, & z \leq z_1,\\
(1+z_1)^{n_1-n_2} \ (1+z)^{n_2}, & z>z_1 \end{array} \right.
\eeq 
$R_{\rm GRB}(z)$ is the comoving rate $dN/dV_{\rm comov} dt_{\rm src}$, where $dV_{\rm comov}$ is the comoving volume and 
$dt_{\rm src}$ is the time interval in the source frame.  
$R_{\rm GRB}(z)$ can be converted to the observed rate $R_{\rm GRB;dz} = dN/d\Omega dz dt_{\rm obs}$ 
in unit of number per solid angle ($d\Omega$), per redshift interval ($dz$), 
and per time interval in the observed frame ($dt_{\rm obs})$ by
\beq
\label{eq:GRBrate_dz}
R_{\rm GRB;dz}(z) = R_{\rm GRB}(z) \ \frac{dV_{\rm comov} dt_{\rm src}}{d\Omega dz dt_{\rm obs}} = \frac{R_{\rm GRB}(z)}{(1+z)} \ \frac{dV_{\rm comov}}{d\Omega dz}
\eeq

\item GRB luminosity function: Again, we adopt the functional form in \citet{Wanderman10}, which is also 
a simple broken power law function,
\beq
\label{eq:GRBlum}
\phi(\rm L) = \frac{dN}{dL} =
\ \left \{  \begin{array}{ll}
(\frac{\rm L}{\rm L_{\star}})^{x}, & \rm L < \rm L_{\star},\\
(\frac{\rm L}{\rm L_{\star}})^{y}, & \rm L > \rm L_{\star} \end{array} \right.
\eeq 
where $x$, $y$, and $\rm L_{\star}$ are adjustable parameters.
Note that the luminosity $L$ refers to the peak luminosity,
not the average luminosity.

\item GRB pulse shape:
Using the observed GRB light curves from the real BAT-detected GRBs with known redshifts, 
we creat a library of rest-frame GRB light curves that contains 139 different GRB pulse shapes.
Data with signal-to-noise ratio higher than $3 \sigma$ are considered as part of the GRB pulses,
while data with signal-to-noise ratio below $3 \sigma$ are considered as noise and thus are ignored 
when constructing the GRB pulse shapes.
The light curves range from duration $T_{90} =$ 2.16 sec to 658.2 sec.
We randomly choose the GRB pulse shape from this library to create simulated light curves in the rest frame. 

\end{enumerate}

\subsection{Simulating the observed light curves with accurate energy response of the BAT}
\label{sect:obs_lc}

We convert the GRB light curve in the rest frame created in part 1 (Sect.~\ref{sect:lc_rest}) into the observational light curve 
in units of photon count rate.
The conversion is calculated using XSPEC\footnote{http://heasarc.nasa.gov/xanadu/xspec/}, 
a program that is capable of converting flux into photon count rate 
based on the energy response of the instrument with an input GRB spectrum \citep{XSPEC}. 
We adopt the commonly used Band function as the GRB spectrum \citep{Band93}, which has the form
\beq
\label{eq:band}
\frac{dN}{dE} =\ \left \{  \begin{array}{ll}
K_1 \ E^{\alpha} \ \rm exp[-E(2+\alpha)/E^{\rm obs}_{\rm peak}], & E < \frac{(\alpha - \beta) \ E^{\rm obs}_{\rm peak}}{2+\alpha},\\
K_2 \ E^{\beta}, & E \geq \frac{(\alpha - \beta) \ E^{\rm obs}_{\rm peak}}{2+\alpha} \end{array} \right.
\eeq
$E^{\rm obs}_{\rm peak}$ is the peak energy in the $\nu F_{\nu}$ spectrum in the observer frame.
The normalization factors $K_1$ and $K_2$ is decided by the GRB luminosity after redshifting into the observer frame.
We assign different $\alpha$ and $\beta$ in our mock GRB sample based on the observed spectral distribution reported in \citet{Sakamoto09}. 
We leave the $E^{\rm obs}_{\rm peak}$ distribution to be one of the adjustable functions
to explore possible consequences when different distributions are used.

We also take into account different energy responses of bursts coming from different angles relative to 
the detector plane. In other words, for the same burst, the simulation will create a light curve with 
higher photon count rate if the burst is on-axis, and lower photon count rate if the burst is off-axis.
Moreover, 
BAT separates the detector into four quadrants and can trigger an event 
based on the light curve recorded in only some quadrants of the detector. The reason for doing 
this is to reduce noise for those GRBs coming from off-axis angles that illuminate only part of the detector.
By calculating the signal-to-noise ratio of only the illuminated part of the detector, 
BAT can get a higher signal-to-noise ratio for these bursts and hence 
increase the chance of triggering these off-axis bursts.

\begin{table}[h!]
\caption{\label{tab:grid_id}
Relation of the grid ID and the burst incoming angle (in degrees) relative to the detector's plane, with zero degrees indicating an on-axis burst. There are no grid ID 6 and 28.
}
\begin{center}
\begin{tabular}{||c|c||c|c||c|c||c|c||c|c||}
\hline\hline
Grid ID & Angle ($^{\circ}$) & Grid ID & Angle ($^{\circ}$) & Grid ID & Angle ($^{\circ}$) & Grid ID & Angle ($^{\circ}$) \\
\hline
1 & 50.69 & 10 & 19.27  & 18 & 26.56 & 26 & 46.63 \\
2 & 40.68 & 11 & 31.39  & 19 & 44.99 & 27 & 56.99 \\
3 & 35.02 & 12 & 46.63  & 20 & 56.29 & 29 & 50.66 \\
4 & 40.68 & 13 & 56.99  & 21 & 56.99 & 30 & 40.68 \\
5 & 50.66 & 14 & 56.29  & 22 & 46.63 & 31 & 35.02 \\
7 & 56.99 & 15 & 44.99  & 23 & 31.39 & 32 & 40.68 \\
8 & 46.63 & 16 & 26.56  & 24 & 19.27 & 33 & 50.66 \\
9 & 31.39 & 17 & 0.076  & 25 & 31.39 & & \\

\hline\hline
\end{tabular}
\end{center}
\end{table}

\begin{figure}[!h]
\begin{center}
\includegraphics[width=0.65\textwidth]{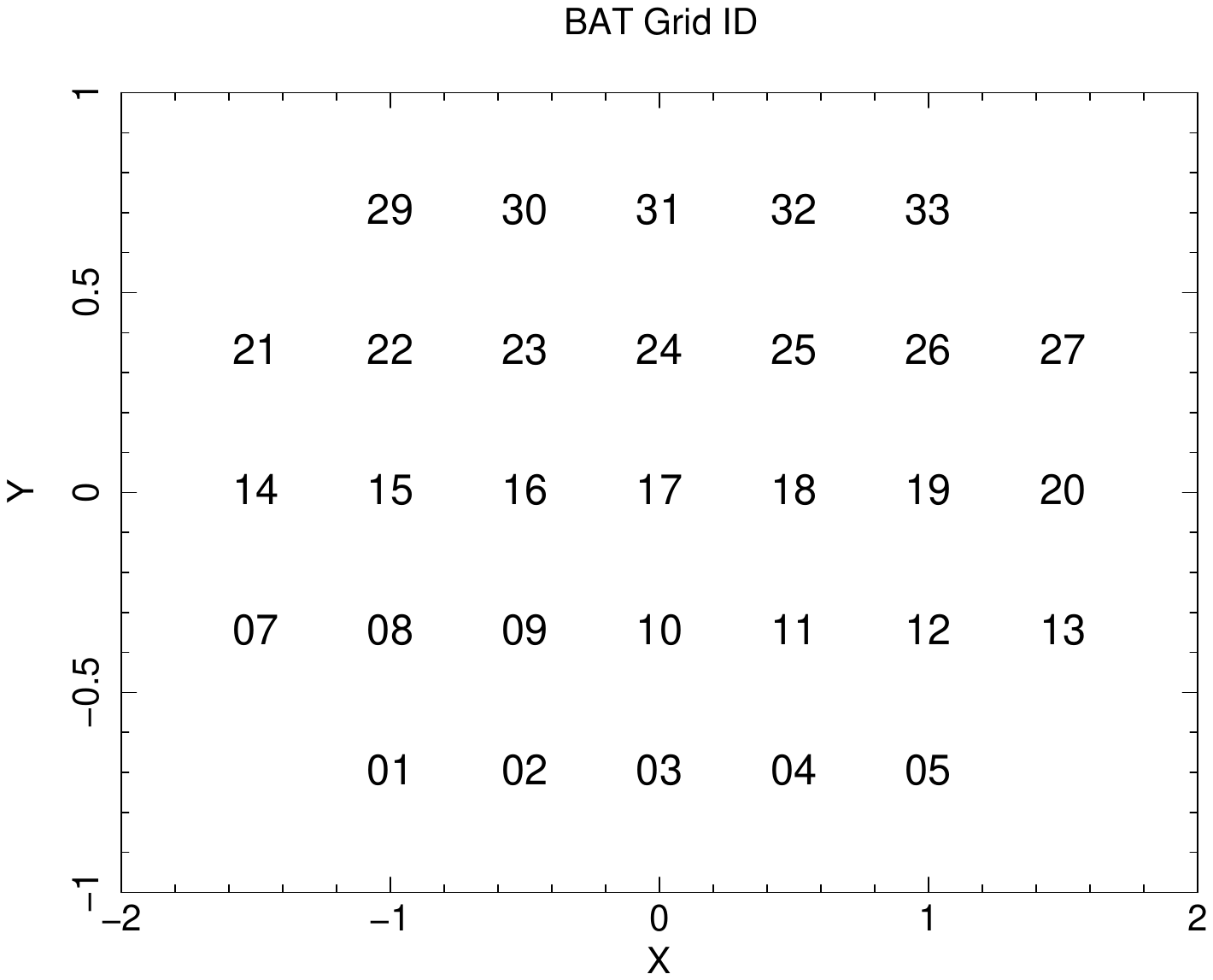}
\end{center}
\vspace{-20pt}
\caption{
Location of the Grid ID on the detector plane. X and Y are the detector coordinate with arbitrary unit. 
The center of the detector is located at X=0, Y=0. 
}
\label{fig:grid_id}
\end{figure}

\begin{figure}[!h]
\begin{center}
\includegraphics[width=0.65\textwidth]{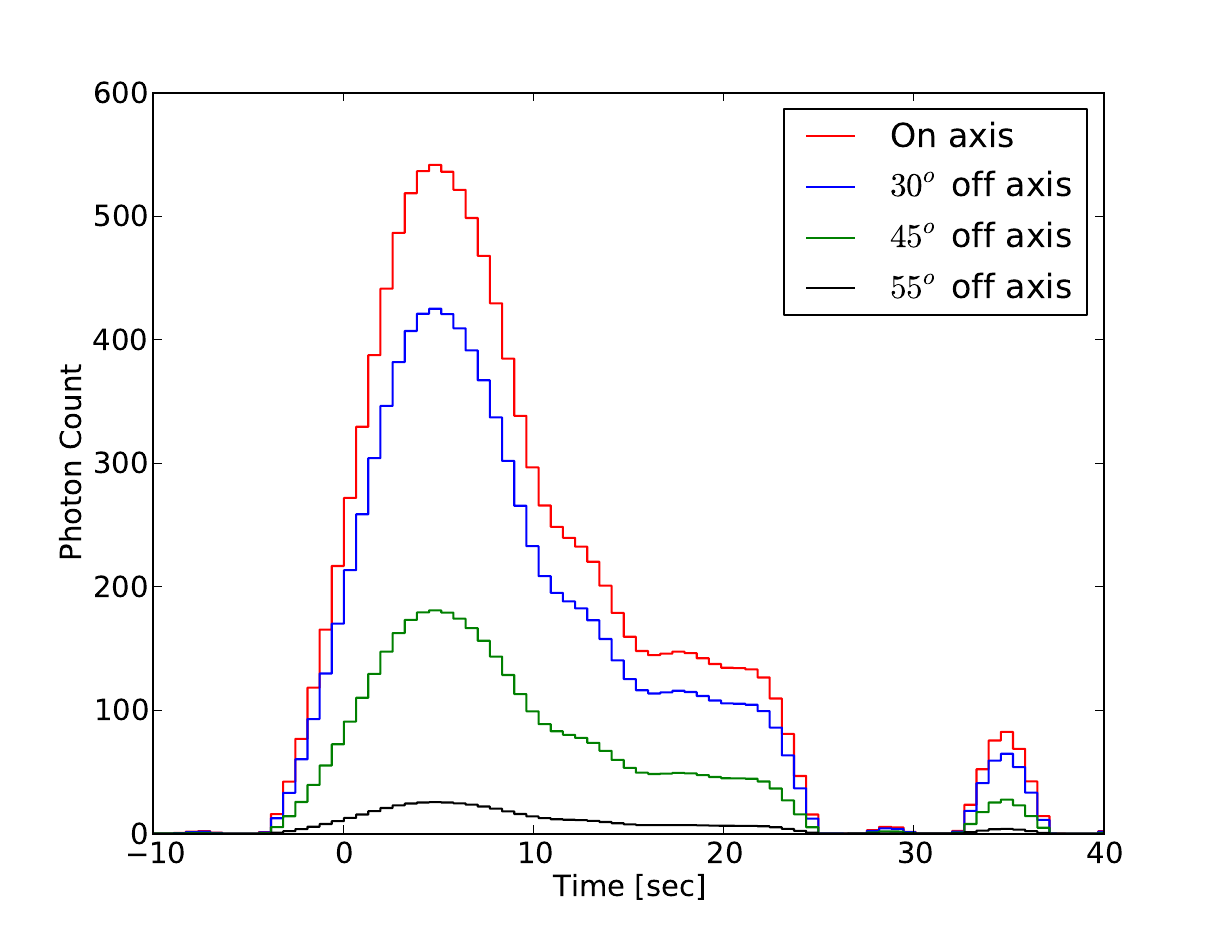}
\end{center}
\vspace{-20pt}
\caption{
Simulated GRB light curves of the same burst with different incident angle relative to the detector plane. 
Specifically, bursts in the figure with incident angles $= \{0^{o}, 30^{o}, 45^{o}, 55^{o}\}$
are generate from bursts with grid ID = $\{17, 16, 15, 14\}$, respectively.
Light curves are binned in 0.64 second.
}
\label{fig:GRB_lc_gridid}
\end{figure}

We label the BAT detector plan with some grid ID numbers.
GRBs with different incident angles will fall on different locations on the detector plane 
with different ``grid ID'', and different calibrations are used based on the ``grid ID'' of the burst.
Figure~\ref{fig:grid_id} plots the relative location of each grid ID on the detector plane.
Table~\ref{tab:grid_id} shows how the incident angle corresponds to each grid ID.
An incoming angle of zero degrees corresponds to an on-axis burst, that is, the burst 
is directly above the detector plane. 
A larger incoming angle indicates that the burst is more off-axis.
Grid ID 6 and 28 do not exist due to historical naming reasons.
To simulate the actual trigger algorithm, 
our program also simulates the partial illumination factor based on the ``grid ID'' (i.e., incident angle) of the burst, 
and creates light curves for four different quadrants of the detector accordingly.
Figure~\ref{fig:GRB_lc_gridid} shows an example of simulated GRB light curves of the same burst with 
different grid IDs. One can see from the figure that 
the incident angle has a significant effect on the detector sensitivity. 
An off-axis burst can be $\sim 20$ times dimmer than an on-axis burst,
and hence become undetectable.

\begin{figure}[!h]
\begin{center}
\includegraphics[width=0.7\textwidth]{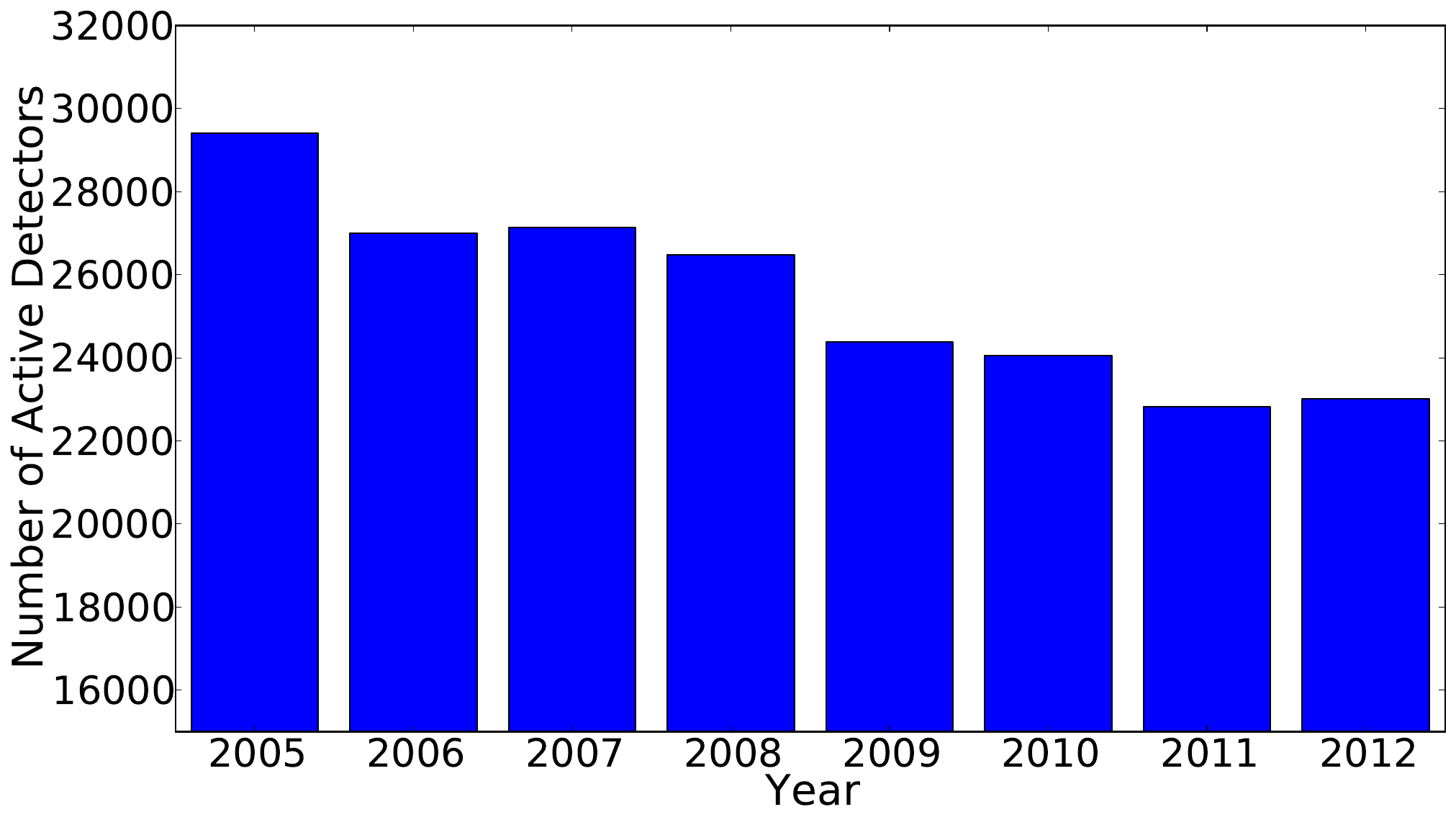}
\end{center}
\vspace{-20pt}
\caption{
The average number of BAT's active detector as a function of year. 
}
\label{fig:ndet}
\end{figure}

In addition, the active number of BAT's detector changes with time. 
BAT has a total number of 32768 detectors. However, during the eight years
of operation, many detectors are getting noisier and are automatically or manually turned off.
Therefore, the average number of active detectors decreases each year, as shown in Fig.~\ref{fig:ndet}.
\begin{figure}[!h]
\begin{center}
\includegraphics[width=0.65\textwidth]{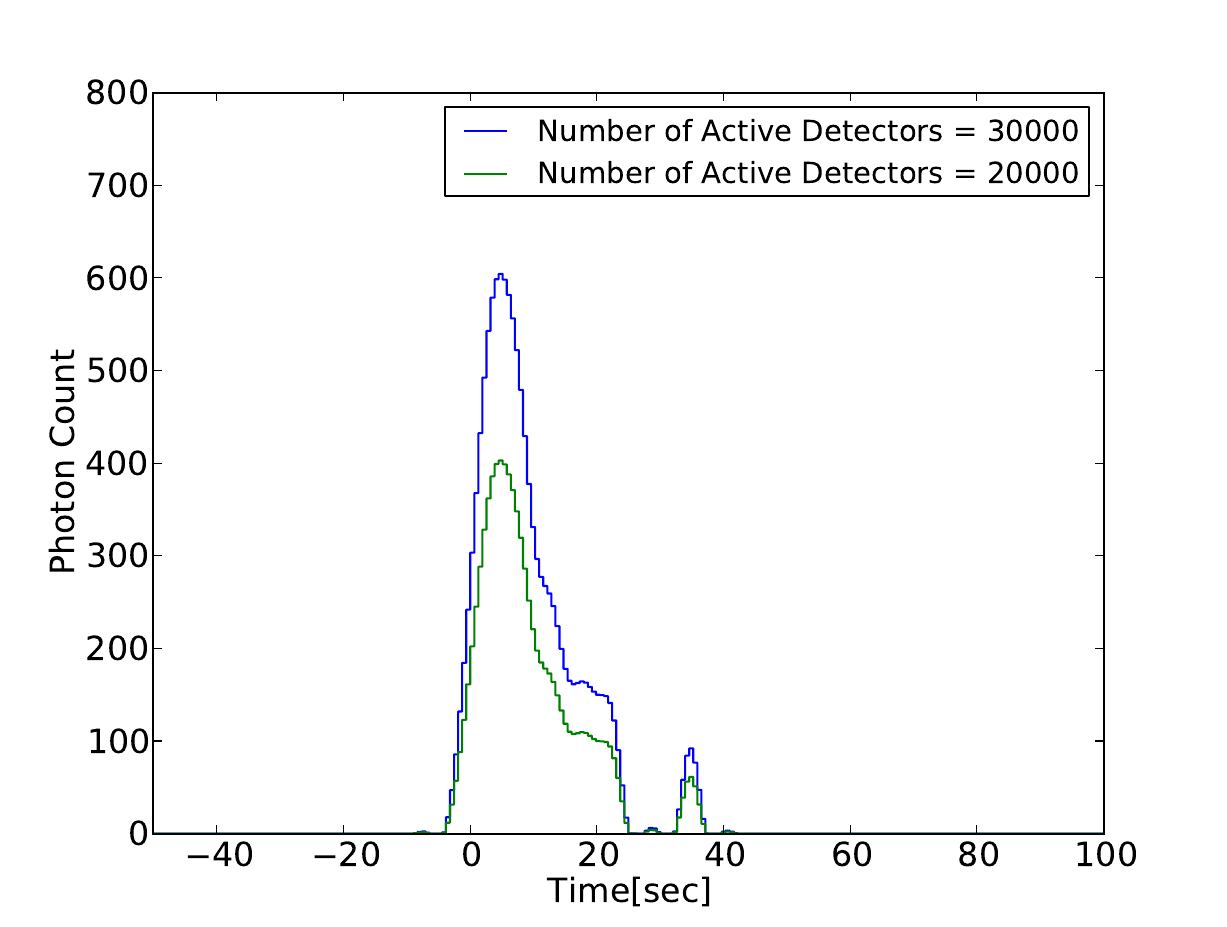}
\end{center}
\vspace{-20pt}
\caption{
Simulated GRB light curves of the same burst with different number of active detectors. 
Light curves are binned in 0.64 second.
}
\label{fig:ndet_GRB}
\end{figure}
This factor is taken into account in our program by simulating light curves 
with different energy responses according to different number of active detectors.
A burst would have a simulated light curve with higher photon count rates 
if we set a larger number of active detectors. Fig.~\ref{fig:ndet_GRB} shows an example 
of how the simulated light curve changes with different number of active detectors.

\begin{figure}[!h]
\begin{center}
\includegraphics[width=0.8\textwidth]{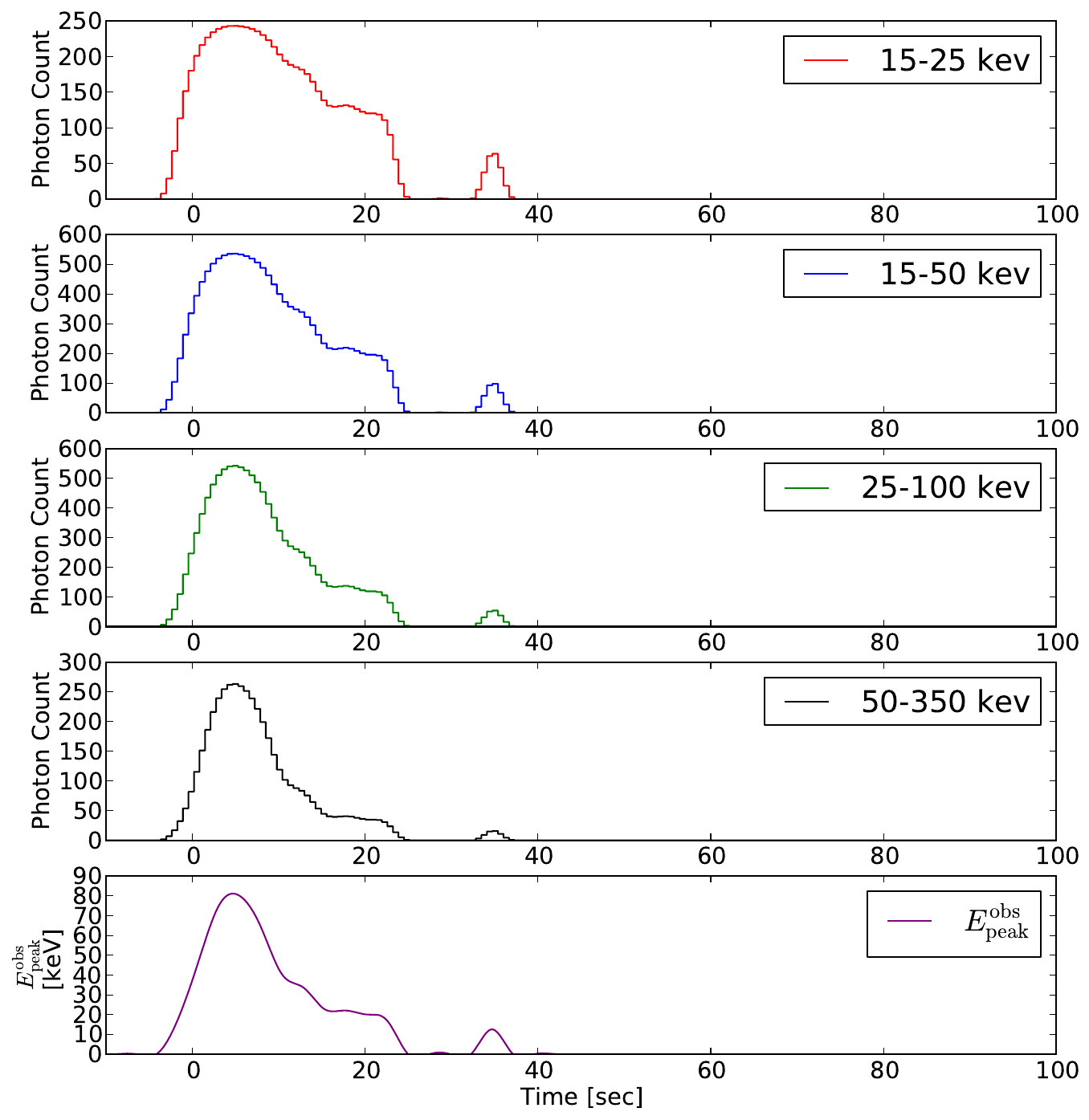}
\end{center}
\caption{
An example of the simulated GRB light curve with spectral ($E^{\rm obs}_{\rm peak}$) evolution. Light curves are binned in 0.64 second.
}
\label{fig:GRB_epeak_evo}
\end{figure}

Furthermore, many observed GRBs show
time-evolving spectra. 
Therefore, we implement the option to include spectral evolution in our simulation.
A majority of well-studied GRBs 
show a hard-to-soft spectral evolution as bursts become dimmer \citep[e.g.,][]{Liang96, Crider99, Ryde00, Zhang07, Racusin08, Starling08, Yonetoku08, Page09,Filgas11}.
Some studies proposed that the spectral evolution is related to the pulse shape \citep{Liang96, Crider99,Ryde00}. 
However, there are also bursts that do not show strong spectral evolution nor a monotonic behavior \citep{Zhang07}.
For simplicity, we adopt the assumption that $E^{\rm obs}_{\rm peak}$ evolves with the pulse shape.
In other words, $E^{\rm obs}_{\rm peak}$ increases as a GRB becomes brighter, and decreases as the burst fades.
Fig.~\ref{fig:GRB_epeak_evo} demonstrates an example of adding $E^{\rm obs}_{\rm peak}$
evolution in the simulated light curve.

\begin{figure}[!h]
\begin{center}
\includegraphics[width=0.9\textwidth]{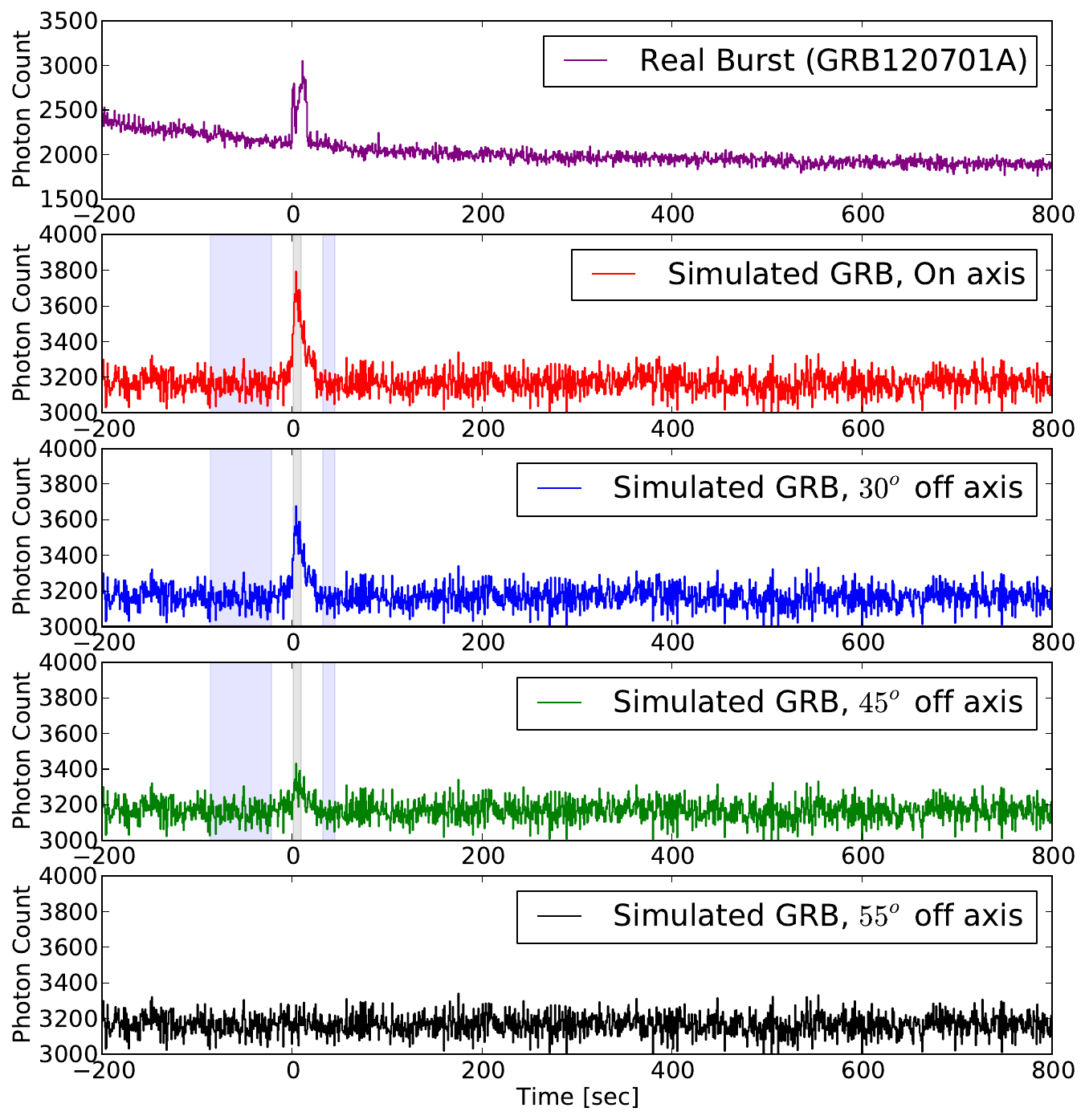}
\end{center}
\caption{
An example of the simulated GRB light curve with background noise. For comparison, a real GRB light curve is also shown in
the top panel. Simulated light curves with different incident angles are plotted in different panels. These are the same light curves shown in
Fig.~\ref{fig:GRB_lc_gridid}, but now with background noise added. 
The grey and blue shaded areas on top of the simulated light curves indicate the foreground and background periods 
that give the highest signal-to-noise ratio using our trigger simulator (see Sect.~\ref{sect:rate_trigger}). 
Light curves are binned in 0.64 second.
}
\label{fig:GRB_lc_sim}
\end{figure}

Finally, we add some background noise to the simulated GRB light curves. 
We create a library of 371 background levels from the light curves of real BAT-detected GRBs. 
The backgrounds added to the simulations are randomly chosen from this library
and are fluctuated with Gaussian noise.
The lower four panels of Fig.~\ref{fig:GRB_lc_sim} show several examples of the simulated light curves with background noise 
in the observational frame with units of photon count rate. 
The light curves shown here are in the 25-100 keV energy band, and are the summaztions of all four detector quadrants.
Each panel shows the same burst with different incident angles, corresponding to the original light curve in Fig.~\ref{fig:GRB_lc_gridid}
(without background noise). The example shows that the $55^{o}$ off-axis burst is now completely buried under the background noise,
and thus is undetectable.
The grey and blue shaded regions indicate the foreground and background periods from the trigger criterion that detects the burst 
in our simulation (see the following section, Sect~\ref{sect:rate_trigger} for more discussion about the trigger simulator). 
For comparison, the top panel of Fig.~\ref{fig:GRB_lc_sim} shows an example of a real GRB light curve (GRB120701A).
The real GRB light curve is also a summation of light curves from all four detector quadrants, and in the energy band 25-100 keV.
Note that real GRB light curves might not have flat backgrounds like those in our simulations, as shown in this example.

\subsection{BAT-trigger simulator that simulates the complex trigger algorithm of the BAT}

The code we developed is capable of simulating both the rate trigger and image trigger processes. 
The following sections describe in detail the methods we adopt to simulate the BAT-trigger algorithm.

\subsubsection{Simulating the first part of the trigger process: the rate trigger}
\label{sect:rate_trigger}

Our program follows the same procedure adopted by BAT for the rate trigger. That is,
we move through the light curve and calculate the signal-to-noise ratio of each bracketed time period ($\tau_{\rm bracket}$)
specified by a rate trigger criterion.
Each bracketed time period contains one foreground period, one or two background periods
(depending on the trigger criterion), and an elapse time period between 
the foreground and background periods.
For those trigger criteria that contain two background periods,
the background periods $\tau_{\rm b1}$ and $\tau_{\rm b2}$ are placed before and after the foreground period $\tau_{\rm f}$, respectively.
Both of the background periods are separated from the foreground period by the elapse period $\tau_{\rm e}$,
to make sure the background light curve does not include photon counts from the source.
For the trigger criteria that only use one background period $\tau_{\rm b1}$,
it is placed before the foreground period.
An elapse period $\tau_{\rm e}$ is also placed between 
$\tau_{\rm b1}$ and $\tau_{\rm f}$ for these one-background criteria.

We follow \citet{Graziani03} in defining the signal-to-noise ratio,
\begin{align}
&\mu_{\rm f} = \frac{\tau_{\rm b1} + \tau_{\rm b2}}{\tau_{\rm f}} \ \Sigma^2, \\
&\Sigma^2 = \tau^2_{\rm f} \ \left( \frac{\tau^2_{\rm b1}}{n_{\rm b1}} + \frac{\tau^2_{\rm b2}}{n_{\rm b2}} \right)^{-1}, \\
&\mbox{Signal-to-Noise Ratio} = \frac{n_{\rm f} - \mu_{\rm f}}{(\mu_{\rm f} + \Sigma^2)^{1/2}}. 
\label{eq:snr}
\end{align}
The signal-to-noise ratio is calculated by Eq.~\ref{eq:snr},
using all the time periods described above 
and the corresponding photon counts in those periods. 
In the equations, $n_{b1}$, $n_{b2}$, and $n_{\rm f}$ represent the photon counts in $\tau_{\rm b1}$, $\tau_{\rm b2}$, and $\tau_{\rm f}$, respectively. 
For the criteria with only one background, the signal-to-noise ratio is calculated by this same equation, 
with all the terms related to the second background (i.e., $\tau_{\rm b2}$, $n_{b2}$) dropped. 

After the signal-to-noise ratio is calculated, the program moves to the next time step, which in our simulation is set to be the next light curve bin,
and calculates the signal-to-noise ratio in that time period. The program moves through the whole light curve and records
the time periods where the signal-to-noise ratios are higher than the threshold determined by the specific trigger criterion.

Each rate trigger criterion has different $\tau_{\rm b1}$, $\tau_{\rm b2}$, $\tau_{\rm f}$, $\tau_{\rm e}$, and signal-to-noise ratio threshold for triggering. 
Additionally, each trigger criterion uses light curves from different energy bands, and light curves from different summations
of the four detector quadrants. The program thus goes through the light curves multiple times with settings specified by each criterion. 
The trigger criteria we adopt are identical to those used by the BAT flight software.


\subsubsection{Mimicking the second part of trigger process: the image threshold}

After an event is triggered by one or more rate trigger criteria, BAT creates real sky images for further confirmation and localization.
However, creating images is very time-consuming and requires substantial computational power. 
Fortunately, it is possible to mimic the image threshold without creating real images in our simulations.
Instead, we use information from the light curves and results of the rate triggers.
This method allows us to calculate the image threshold without high demands of computational resources,
and thus to be able to 
explore a larger parameter space in our search of the GRB characteristics.

This approximation is feasible in our simulation because of the following reasons.
For a real burst observation, images are created to obtain better localization information and clarify 
that the event is not from a known stellar object. However, in our simulation all the sources generated are 
GRBs. Hence, there is no confusion with other sources and all we need to know is whether the 
burst is triggered or not.
Another reason for BAT to create a real image is to determine the real background level at
the time when the burst happens. In reality, the background is constantly changing with time.
Therefore, the background level when the burst happens is likely to be different than the background level before 
or after the burst. However, we assume a flat background with Gaussian noise in our simulation,
and thus the background level should be time independent and the photon count rate calculated from the 
background period in the rate trigger should be the same as that obtain by an real image.
Therefore, we can mimic the image threshold based on information gathered from the rate trigger. 

When a rate trigger criterion successfully triggers an event, 
BAT uses the accumulated photon counts in the foreground period of that rate trigger criterion
to create an image. The photon counts used to create this image are added from all four quadrants of the detector,
even if the rate trigger criterion that triggered this event only uses light curves from part of the detector.  
Therefore, when mimicking the image threshold, we use the foreground and background periods from the successful rate trigger criterion,
but use the light curves from all four quadrants to calculate the signal-to-noise ratio.

Determining the corresponding signal-to-noise ratio of the image threshold using only information from the light curve can be a little bit complicated.
In reality, the image signal-to-noise ratio is calculated from the real image by fitting a point spread function.
Therefore, the signal-to-noise ratio might not be identical to the signal-to-noise ratio calculated using
light curves by Eq.~\ref{eq:snr}. 
In an ideal case of flat background and only one source (i.e., the burst) per image, 
the signal-to-noise ratio calculated from point spread function fitting should have 
a one-to-one correlation to the signal-to-noise ratio calculated from the light curve.
Thus, we can approximate the image threshold if we can find 
the corresponding signal-to-noise ratio calculated from a light curve to that estimated 
from an image. That is, we need to know how to convert the image signal-to-noise threshold
to a threshold in the light curve domain.

\begin{figure}[!h]
\begin{center}
\includegraphics[width=0.7\textwidth]{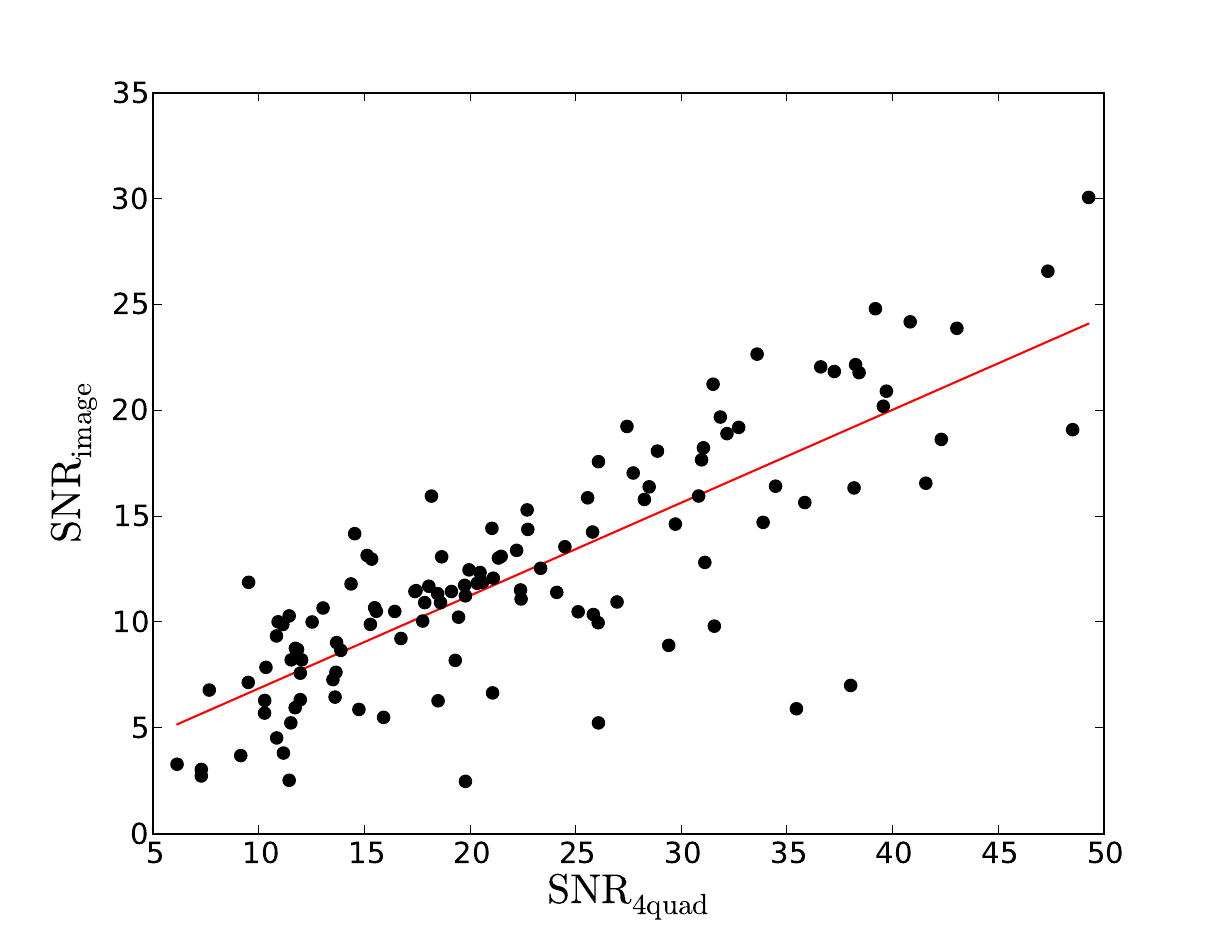}
\end{center}
\caption{
Correlation of $\rm SNR_{\rm image}$, the image signal-to-noise ratio using the BAT software (HEASoft), and $\rm SNR_{\rm 4quad}$, the signal-to-noise ratio from the four quadrant light curves.
Red line shows the $\chi^2$ fitting, which has the form of $\rm SNR_{\rm image} = 2.47 + 0.44 \ \rm SNR_{\rm 4quad}$.
}
\label{fig:image_snr}
\end{figure}

To determine the corresponding signal-to-noise ratio of the image threshold,
we apply the BAT-trigger simulator to 121 light curves of real BAT-detected GRBs.
Using the foreground and background periods of the successful trigger criteria,
we calculate the signal-to-noise ratio using light curves cladded from all four quadrants.
Independently, we create real images for these real GRBs using the foreground period
determined by the successful trigger criteria and calculate the image signal-to-noise ratio
using the ground software of BAT (HEASoft\footnote{http://heasarc.nasa.gov/lheasoft/}).
We plot the signal-to-noise ratios calculated by these two methods to find the correlation in Fig.~\ref{fig:image_snr}.
As discussed earlier,
the scatter of the points shown in Fig.~\ref{fig:image_snr} is likely due to 
the large fluctuation of the backgrounds of these real GRB light curves,
and/or from contamination of other sources in the light curves and images.
For a first-order approximation, we simply fit a linear function to the correlation of the image signal-to-noise ratio 
calculated from the BAT software HEASoft ($\rm SNR_{\rm image}$)
and the signal-to-noise ratio calculated from photon count rate using all four quadrant light curves ($\rm SNR_{\rm 4quad}$).
This fitted line has the functional form of $\rm SNR_{\rm image} = 2.47 + 0.44 \ \rm SNR_{\rm 4quad}$,
with reduced $\chi^2$ $=11.12$ (degree-of-freedom = 118).
The signal-to-noise ratio for the real image threshold is $\sim 6.5$ to 7.0, which corresponds to a signal-to-noise ratio of $\sim 10$
calculated from light curves. Thus, we adopt the signal-to-noise ratio = 10 to be our threshold for mimicking the image threshold.

\subsubsection{The image trigger}

In addition to the regular two-stage trigger algorithm (rate trigger followed by image threshold),
BAT also creates images regularly in the $15-50$ keV energy range for longer durations ($\gtrsim 1$ minute) to search for bursts missed by the rate triggers.
GRBs found in this independent image trigger process usually have low fluxes but high fluences, 
which are likely due to relatively long and slow-changing light curves.
These GRBs are thus hard to detect using rate triggers with shorter trigger durations.

The signal-to-noise threshold for this image trigger process is $\sim 7.0$ to 7.5, 
which is similar to the
image threshold following the rate trigger.
Therefore, we adopt the same method as discussed in the previous section with the same criterion of signal-to-noise ratio $\sim 10$
for mimicking the image trigger.
However, we modify the foreground and background durations
to replicate the longer exposure time in each image.

As mentioned in Sect.~\ref{sect:trigger_algorithm},
the BAT flight software makes images with many different 
durations, with some durations only available when 
a rate trigger is active,
and thus gives some randomness to
the ranges of exposure time of these real images.
In order to determine the foreground durations for the image trigger process in our simulation,
we list the time intervals of the image exposure times for the real BAT image-triggered GRBs,
and adopt them as the foreground durations in our simulation.
These durations are 64, 72, 88, 120, 128, 168, 192, and 320 seconds.
The background durations are set to be the same as the foreground durations in our simulation to calculate 
the signal-to-noise ratio of the bursts.
We put in a 32 second elapse time between the background and foreground periods, 
to make sure the background is not contaminated by the burst light curve 
when there is a detection in the simulation.
We only consider triggers in the $15-50$ keV band for the image trigger,
because this is the only energy band used by BAT during this process.

\subsubsection{Comparing our simulation with BAT's sensitivity}
\label{sect:compare_sensitivity}

To test whether our program correctly simulates the complex BAT-trigger algorithm, 
we compare the GRB peak fluxes of the ``triggered'' bursts in our simulation 
to those measured from the real GRBs detected by BAT.

\begin{figure}[!h]
\begin{center}
\includegraphics[width=0.6\textwidth]{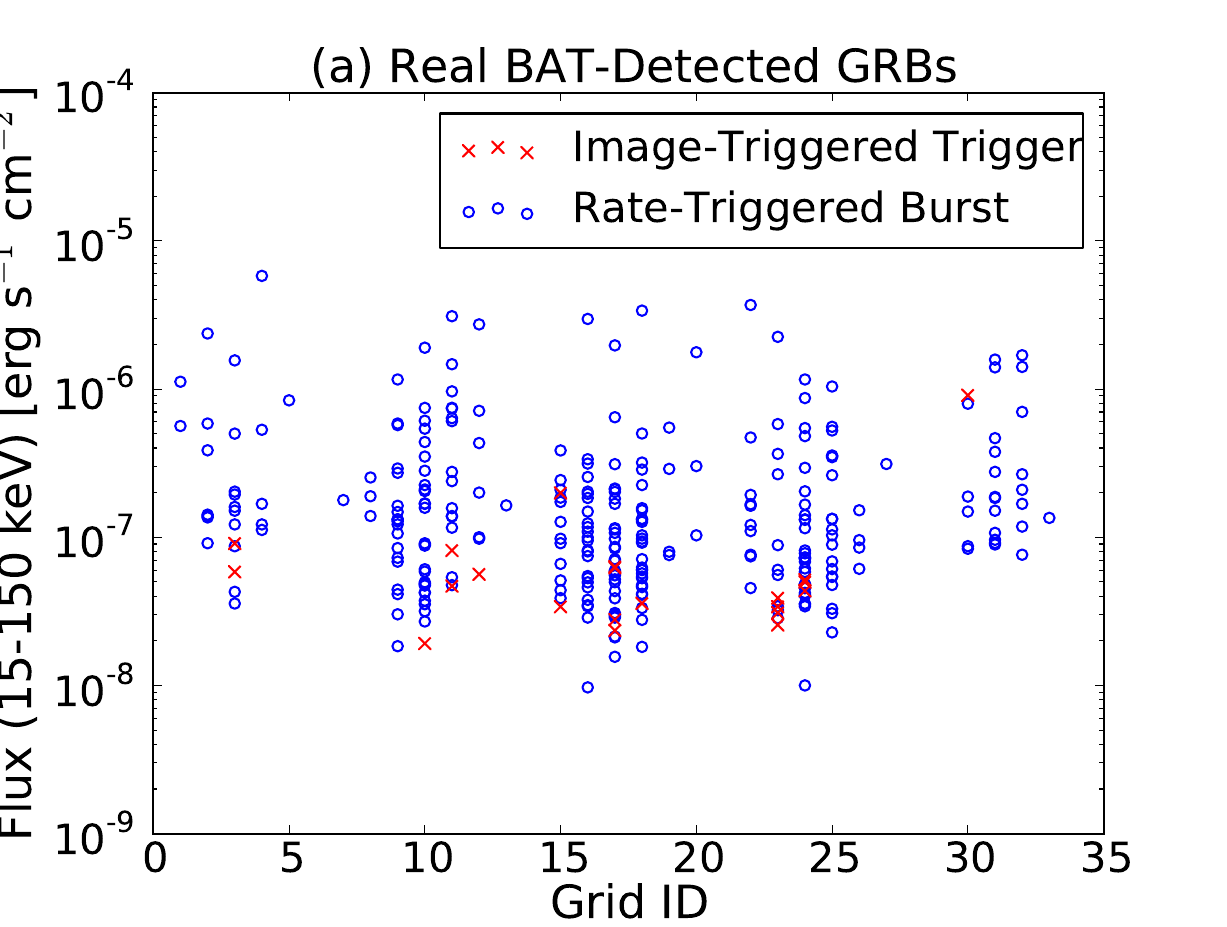}
\includegraphics[width=1.0\textwidth]{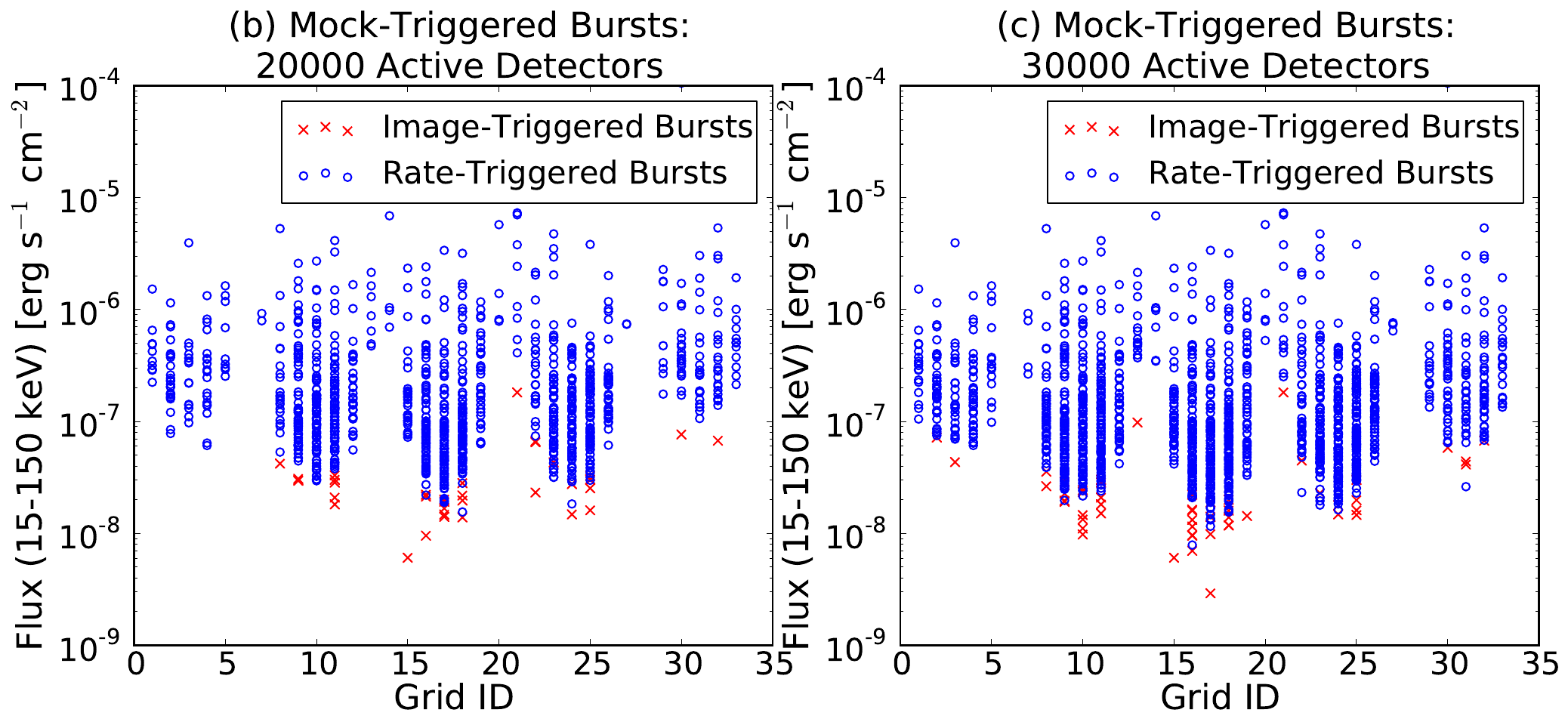}
\end{center}
\caption{
{\it Panel (a)}: 1-s peak energy flux vs. grid ID for the BAT detected bursts from GRB041223 to GRB091221 \citep{Sakamoto11}. 
{\it Panel (b)}: Peak energy flux vs. grid ID for the triggered GRBs in the mock sample, assuming 20000 active detectors.
{\it Panel (c)}: Peak energy flux vs. grid ID for the triggered GRBs in the mock sample, assuming 30000 active detectors.
}
\label{fig:sensitivity_grid}
\end{figure}

Panel (a) of Fig.~\ref{fig:sensitivity_grid} shows the peak fluxes of real BAT-detected GRBs with respect to 
the grid ID of the detector plane. As described in Sect~\ref{sect:obs_lc}, 
the grid IDs are simply the number labels on the detector plane, and thus correspond to 
the incoming angles of the bursts relative to the normal axis of the detector.
Each point in the figure represents one burst.
The fluxes of these bursts are adopted from \citet{Sakamoto11}.
Blue dots in the plot indicate GRBs detected by rate triggers, while red crosses represent bursts found by image triggers.

Panels (b) and (c) of Fig.~\ref{fig:sensitivity_grid} plot the fluxes from the mock ``triggered'' bursts in our 
simulations with 20000 and 30000 active detectors, respectively. 
Again, blue dots show the rate-triggered bursts, while red crosses indicate the image-triggered events.
Because the number of active detectors decreases with time, 
the sensitivities using two extreme numbers of active detectors are plotted for comparison.
Results show that decreasing the number of active detectors from 30000 to 20000 
reduces the sensitivity by a factor of $\sim 3$,
and has more significant impact on off-axis bursts than on-axis events.
Moreover, the sensitivity of the image trigger is less affected by 
reducing the number of active detectors. 
The input GRB characteristics of this mock sample are based on the best result in our search, which is summarized in Table~\ref{tab:best_result}, Sect.~\ref{sect:rate}. 
This sample creates plenty of bursts that have fluxes in the range of $10^{-9} \  \rm erg \ s^{-1} \ cm^{-2}$ to $10^{-6} \ \rm erg \ s^{-1} \ cm^{-2}$.
Therefore, the non-detection of low-flux bursts is due to the sensitivity of our trigger-simulation program, 
instead of the lack of low-flux events.

Results show that the rate trigger process in our trigger simulator can detect 
GRBs with fluxes $\sim 10^{-8} \ \rm erg \ s^{-1} \ cm^{-2}$ for directly on-axis bursts, and fluxes $\sim 10^{-7} \ \rm erg \ s^{-1} \ cm^{-2}$ for extremely off-axis events,
which is very similar to the real BAT sensitivity.
It is harder to compare the sensitivity of the image trigger process, due to the very low number of statistics in real BAT detections. 
However, in general the image trigger process in our simulations detects bursts with fluxes that are slightly lower than those
found by the rate trigger algorithm, as expected. A similar trend is also seen in the real BAT-detected GRBs.



\section{Observational distributions of the GRB characteristics}
\label{sect:BAT_observables}

In our search for the intrinsic GRB characteristics, we need 
to compare the GRB properties from our mock-triggered sample,
which contains the simulated bursts that are triggered by our trigger simulator,
to the GRB properties from the real BAT-detected bursts.
In this section, we describe the observational GRB samples we adopt
to be compared with our simulated bursts. We also discuss the challenges
and uncertainties in these observational measurements. 

The main GRB characteristics we use for comparison are
the redshift and the peak-flux distributions. 
We iteratively adjust the input parameters until the simulated results match well with
both the observed redshift and peak-flux distributions.
The peak-flux distribution is chosen as one of the main guides in our search because the peak flux can be measured directly from
observations and thus is less uncertain, as described in Sect.~\ref{sect:BAT_peakflux}.
The redshift distribution is also selected
because our main goal here is to find the intrinsic GRB rate. 
In addition, we also consider $E^{\rm obs}_{\rm peak}$ and $E_{\rm iso}$ from real observations.
However, due to the difficulties in the measurements of $E^{\rm obs}_{\rm peak}$ and $E_{\rm iso}$ and their large uncertainties
in the BAT data, 
these two properties are only used as references but not as the primary guides 
in our search. 

In all these observed GRB samples we adopted, 
the bursts found by ground analysis, instead of flight software,
are removed. This is because our trigger simulation follows the 
method of the flight software. Moreover, currently the ground analysis
is mainly done by human and hence is less systematic.
There are only a few ground-detected bursts in the observed GRB samples
(zero in the redshift sample; 7 out of 409 in the flux sample; 9 out of 423 in the $E_{\rm peak}$ sample; see Sect~\ref{sect:BAT_redshift} to Sect~\ref{sect:BAT_Eiso}).
Therefore, excluding these bursts does not have significant effect on the results.

\subsection{The redshift distribution of the BAT-detected GRBs}
\label{sect:BAT_redshift}

Around $30\%$ of all the BAT-detected GRBs have measured redshifts \citep{Gehrels12}.
These redshifts are usually measured from the burst afterglows and/or host galaxies
observed by follow-up ground observations.
Therefore, the GRB redshift sample suffers greatly from observational biases,
and is highly skewed toward lower redshifts.
As there are many unpredictable factors that affect the redshift measurements,
such as the local weather conditions and the number of telescopes that are available for follow-up observations, 
it is difficult to construct a complete redshift sample of the observed GRBs.

Despite the complications, many studies have discussed the observational selection biases and their effect on 
the GRB redshift distribution, and explored possibilities to construct a complete GRB sample \citep[e.g.,][]{Coward08, Fynbo09, Jakobsson12, Hjorth12, Salvaterra12, Coward13}.
In particular, \citet{Fynbo09} carefully consider possible observational constraints 
on performing follow-up observations and create a GRB sample that is less affected 
by the observational biases.
These authors set up a number of criteria to 
select GRBs that have optimal conditions for follow-up observations.
For example, bursts that are too close to the Sun or the Moon, or have high Galactic extinction, are removed from the sample,
because it is harder to perform follow-up observations for these bursts 
and thus these bursts need to be extremely bright to have measured redshifts.

There are 79 GRBs with redshift measurements (either from afterglows, host galaxies, or both)
in the statistical GRB sample compiled by \citet{Fynbo09} (Table 73 in their paper).
These authors note that
bursts with redshift measurements from the optical afterglows
are not representative for all the {\it Swift} bursts in their statistical sample. 
The GRBs without optical spectroscopy in their sample
are likely to suffer from dust obscuration other than the Galactic extinction, 
such as dust in the GRB host galaxies.
Similar conclusion has also been obtained by
The Optically Unbiased Gamma-Ray Burst Host (TOUGH) survey,
which performs a systematic search for the GRB host galaxies
and found that
the host galaxies of GRBs with no optical and/or near infrared 
afterglows are significantly brighter and redder than 
those with optical and/or near infrared 
afterglows \citep{Hjorth12}.
Although it is possible to use host galaxy detections to recover these missing redshifts of 
those bursts without optical afterglows,
this approach introduces other biases due to
host galaxy brightness and 
the optical survey sensitivity.  

We therefore adopt the GRBs from the statistical sample constructed in \citet{Fynbo09} 
that have redshift measurements either from only afterglows, or from both afterglows and host galaxies. 
We exclude bursts with redshift measurements only from host galaxies, 
because low redshift bursts have a higher probability of having detectable host galaxies.
We also exclude GRBs with photometric redshifts due to their large uncertainties.
Additionally, we remove one burst that does not have $E_{\rm iso}$ information (see Sect.~\ref{sect:BAT_Eiso}).
There are 66 bursts in this redshift sample we adopt.

\subsection{The peak-flux distribution of the BAT-detected GRBs}
\label{sect:BAT_peakflux}

The GRB peak fluxes in the BAT energy range $15-150$ keV
can be measured directly from the BAT observations
using the fewest assumptions about the burst characteristics. 
Therefore, the $15-150$ keV peak flux is the least uncertain among 
all the GRB properties we consider here. 
We adopt the $15-150$ keV peak fluxes of the real BAT-detected GRBs reported in \citet{Sakamoto11};
a total number of 402 bursts are given.

\subsection{The $E^{\rm obs}_{\rm peak}$ distribution of the BAT-detected GRBs}
\label{sect:BAT_Epeak}

Not all of the BAT-detected GRBs have measured $E^{\rm obs}_{\rm peak}$. Therefore, we use the $E^{\rm obs}_{\rm peak}$ estimator 
found in \citet{Sakamoto09} to estimate the burst $E^{\rm obs}_{\rm peak}$ based on the power-law index ($\Gamma$) of the burst spectra 
when fitted by a simple power-law model. The power-law indices are reported in \citet{Sakamoto11}.

The $E^{\rm obs}_{\rm peak}$ estimator in \citet{Sakamoto09} can only calculate $E^{\rm obs}_{\rm peak}$ 
in a limited range of power-law indices ($1.3 \leq \Gamma \leq 2.3$), which
corresponds to $E^{\rm obs}_{\rm peak}$ values inside the detectable energy range of the BAT. 
In the cases where the power-law indices are out of the range,
we use the power-law indices to estimate whether $E^{\rm obs}_{\rm peak}$ lies below or above 
the BAT detectable energy range. 
A small power-law index ($\Gamma < 1.3$) indicates that $E^{\rm obs}_{\rm peak}$ falls above 
the BAT detectable energy range ($E^{\rm obs}_{\rm peak} \gtrsim 150$ keV),
while a large power-law index ($\Gamma > 2.3$) implies that $E^{\rm obs}_{\rm peak}$ 
is lower than the BAT detectable energy range ($E^{\rm obs}_{\rm peak} \lesssim 15$ keV).
There are 414 bursts in the $E^{\rm obs}_{\rm peak}$ sample we adopt, 
in which 26 bursts are below the BAT detectable energy range, 
and 80 bursts are above the BAT detectable energy range.

\subsection{The $E_{\rm iso}$ of the BAT-detected GRBs}
\label{sect:BAT_Eiso} 

The total energy output (i.e., the fluence) $E_{\rm iso}$ of an observed GRB is difficult to measure.
Both the burst spectrum and redshift are required to calculate $E_{\rm iso}$. 
However, most of the bursts are lacking redshift measurements (see Sect~\ref{sect:BAT_redshift})
and/or well-constrained $E^{\rm obs}_{\rm peak}$, leading to a poor characterization of the spectra,
especially when $E^{\rm obs}_{\rm peak}$ lies outside of the BAT energy range.
Therefore, even with a burst that has an observed redshift,
estimating $E_{\rm iso}$ requires one to extrapolate the measured spectrum to an energy out of the detectable 
range, and hence introduces uncertainties.
In addition, an observation might not capture the complete light curve of a burst due to the background noise.
In order words, it is likely that BAT only detects the tip of the burst and thus underestimates the 
total energy output of the event.  

\citet{Butler07, Butler10} report a list of $E_{\rm iso}$ values of the BAT-detected bursts. They estimate the $E_{\rm iso}$ 
in the $T_{90}$ burst duration
and energy range $1 - 10^{4}$ keV.
Due to the difficulties in directly measuring $E_{\rm iso}$, these authors adopt a Bayesian approach 
to estimate the values, with a prior $E^{\rm obs}_{\rm peak}$ distribution following results from the observations of {\it CGRO}/BATSE \citep{Preece00},
the primary GRB instrument prior to {\it Swift}.
\citet{Robertson12} compile a list of BAT-detected GRBs that have measured redshifts and $E_{\rm iso}$ values.
The majority of the $E_{\rm iso}$ values of their list are from \citet{Butler07, Butler10} with a few more from \citet{Sakamoto11}.
In addition, these authors calculate $E_{\rm iso}$ for 29 new bursts from their fluence and redshift values based on numerous references
\citep[see][]{Robertson12}.  
The $E_{\rm iso}$ estimations become more uncertain for bursts with $E^{\rm obs}_{\rm peak}$ lying outside of the BAT detectable energy range,
because it is hard to pinpoint the turnover of the spectrum. 
Therefore, for the $E_{\rm iso}$ sample, we only consider bursts with $E^{\rm obs}_{\rm peak}$ in the BAT energy range
and adopt the corresponding $E_{\rm iso}$ values from the list in \citet{Robertson12}.

\section{Searching for the intrinsic GRB characteristics}
\label{sect:rate}

\subsection{General methods}
\label{sect:general_methods}

We modify the parameters for the intrinsic GRB redshift distribution and luminosity function in 
Eq.~\ref{eq:GRBrate} and Eq.~\ref{eq:GRBlum}
to search for a set of the parameters that generates a mock-triggered burst sample that
matches the best with the observed GRB characteristics.
Since we follow the same functional forms for the redshift and luminosity equations as those in \citet{Wanderman10},
we start our search around numbers reported by those authors.

In our search, we simulate 10000 bursts for each set of parameters in order to have enough mock-triggered bursts
to reduce statistical fluctuations.
Our code was run on computers with 4 Intel quad-core Q9650 processors at 3 GHz on the Scientific Linux release 5.9 (Boron).
Without including $E^{\rm obs}_{\rm peak}$ evolution, our code takes around 5 hours to simulate 10000 bursts
and to run them through the trigger simulator, for light curves with bin size = 1.6 seconds.
If $E^{\rm obs}_{\rm peak}$ evolution is included, the simulation takes $\sim 3$ days for a sample of 10000 bursts
with light curves binned into 1.6 seconds.
If we decrease the light curve bin size (i.e., increase the number of light curve bins for each burst),
the time required for the trigger code to go through the whole light curve range increases as the total number of light curve bins.  

The equations for the redshift distribution and luminosity function we adopt contain seven parameters total (see Eq.~\ref{eq:GRBrate} and Eq.~\ref{eq:GRBlum}).
Therefore, to run a complete Monte Carlo simulation and search through the full parameter space 
is highly demanding of computational power and time.
For example, if we explore a range of ten values for each parameter, we will have $10^{7}$ combinations for the parameter set, 
and hence the full Monte Carlo simulations with burst light curves binned to 1.6 seconds will take $\sim 5700$ years for the simulations to finish. 
One possibility might be adopting the Markov Chain Monte Carlo method. 
However, due to the complexity in the GRB observables we are taking into account 
(e.g., cosmic GRB rate, luminosity function, $E^{\rm obs}_{\rm peak}$ distribution), it is difficult to find a good and efficient algorithm that
guarantees speeding up the process significantly
and converging to the correct answer.
Therefore, instead of searching the full parameter space, 
here we only try to find at least one possible set of parameters that generates a good match
with the observed GRB characteristics.

Among all the long rate trigger criteria of the BAT, 250 criteria have foreground periods longer than two seconds. 
Since we focus only on long bursts in this paper, most of the bursts that are triggered by criteria with foreground periods shorter than two seconds should also 
be triggered by criteria with foreground periods longer than two seconds. 
Therefore, to speed up the search process for the GRB properties, 
we only run through the 250 criteria that are longer than two seconds in our main search.
There could be a few scenarios where the long GRBs are ``long'' by virtue of consisting of 
multiple ``short'' spikes that are well-separated (longer than the foreground period of the trigger criteria).
For these bursts, it is possible that they could only be triggered by criteria with 
foreground durations shorter than two seconds.
Hence, once we find a parameter set that matches well with observations,
we will rerun the sample with all the long trigger criteria
to check whether the results remain a good fit with observations.

Additionally,
our main searches are done without including $E^{\rm obs}_{\rm peak}$ evolution,
and with a fixed $E^{\rm src}_{\rm peak}$ distribution,
for the purpose of speeding up the search process as well. 
To decide what kind of $E^{\rm src}_{\rm peak}$ distribution to use in our main search, 
we perform some test runs at the beginning using several very different functions
for the $E^{\rm src}_{\rm peak}$ distribution (see Sect~\ref{sect:Epeak_dist} for detailed discussion of the choices of functions).
We find that in general, assuming some kind of intrinsic relation between $E^{\rm src}_{\rm peak}$ 
and the burst energy output (e.g., luminosity, $E_{\rm iso}$)
seems to create a better match with observations.
Therefore, in our main searches we adopt an $E^{\rm src}_{\rm peak}$ distribution
that follows the Yonetoku relation
of $E^{\rm src}_{\rm peak}$ and $L_{\rm peak}$  \citep{Yonetoku04} (see more discussion in Sect.~\ref{sect:Epeak_dist}). 
Once we find a sample that matches well with the observations, 
we modify the $E^{\rm src}_{\rm peak}$ distribution using several very different functions
to see whether this would cause significant changes in the results.
Similarly, we also run the same set of parameters with $E^{\rm obs}_{\rm peak}$ evolution
to see how the results might change.  
If the effect of adopting a different $E^{\rm src}_{\rm peak}$ distribution and/or $E^{\rm obs}_{\rm peak}$ evolution
turns out to be significant, 
we adjust the parameters in the redshift and luminosity functions and repeat this process.

Furthermore, as discussed in Sect.~\ref{sect:compare_sensitivity}, the sensitivity of BAT, and hence the sensitivity of our trigger simulator,
is slightly different for different numbers of active detectors. Ideally, to simulate the decrease in BAT's sensitivity, 
we need to run our simulation for different numbers of active detectors 
and take an average of all the results. However, this can easily increase the search time to a point that 
it becomes impractical. Therefore, we start our search with one number of active detectors. 
Usually we start with $\sim 25000$ active detectors, which is the medium number of detectors for the eight years of BAT's operation.
Once we find a set of parameters that matches well with the real observations,
we perform the simulations for several different numbers of active detectors and take the average.
Minor adjustments of the parameters might be needed until this averaged result fits well
with the observational GRB characteristics.

To see how good the mock-triggered sample matches the observed GRB characteristics,
we compare several burst properties of the mock-triggered sample with 
those of the real BAT-detected GRBs. 
As discussed in Sect.~\ref{sect:BAT_observables}, we use the redshift and peak flux distributions as our main guides.
For these two distributions, we perform the KS test to quantify how good the matches are 
between the distributions from the mock-triggered sample 
and those from the real BAT-detected GRBs. 
We modify the parameters in the redshift and luminosity functions 
until the KS test (with uncertainties) gives a significance level above $90\%$.
Additionally, we use the $E^{\rm obs}_{\rm peak}$ distribution and the 
$E^{\rm src}_{\rm peak}$-$E_{\rm iso}$ relation as references.
We try to make these two distributions match as well as possible 
to the observed ones. 
However, due to the large and hard-to-quantify uncertainties in $E^{\rm obs}_{\rm peak}$
and $E_{\rm iso}$ (see discussion in Sect.~\ref{sect:BAT_observables}),
we do not require them to match perfectly, and do not perform 
statistical tests on these two distributions.

\subsection{Results of the best-fit parameters}
\label{sect:best_result}

The parameter set shown in Table~\ref{tab:best_result} contains our best fit parameters
for Eq.\ref{eq:GRBrate} and Eq.~\ref{eq:GRBlum}.
This set of parameters generates mock-triggered bursts that have a redshift distribution and peak-flux distribution
that match the best with those from the real observed GRBs. 
Table~\ref{tab:best_result} also lists the KS test values for both the 
redshift distribution and the peak-flux distribution. Both the number ``D'' in the KS-test and the 
significance level of ``D'' (i.e., the probability of ``D'' larger than the reported value) are given in the Table.
The value ``D''  is the key parameter in the KS test, which indicates the maximum distance between the cumulative distribution of the two
tested samples. Smaller ``D'' implies a better match of the two samples.
All of the real BAT-detected bursts we adopted for comparison are GRBs before 2009. 
Therefore, we run the simulations five times (each time generates 10000 bursts) with a different average number of enabled detectors from year 2005 to 2009.
The results of the simulations shown in the following figures are generated from the total 50000 bursts with 
different numbers of enabled detectors. 

\begin{table}[h]
\caption{\label{tab:best_result}
Summary of the set of parameters that generates results that match the best with the observed GRB characteristics.
}
\begin{center}
\begin{tabular}{|c|c|c|c|}
\hline\hline
$R_{\rm GRB}(z=0) [\rm Gpc^{-3} \ yr^{-1}]$ & $z_1$ & $n_1$ & $n_2$ \\
\hline
0.42 & 3.6 & 2.07 & -0.70 \\
\hline\hline
\end{tabular}
\end{center}

\begin{center}
\begin{tabular}{|c|c|c|c|}
\hline\hline
$L_\star [\rm erg \ s^{-1}]$ & x & y & $E^{\rm src}_{\rm peak}$ distribution \\
\hline
$10^{52.05}$ & -0.65 & -3.00 & Modified Yonetoku Relation (Eq.~\ref{eq:Yonetoku_mod})\\
\hline\hline
\end{tabular}
\end{center}

\begin{center}
\begin{tabular}{|c|c|c|c|c|c|}
\hline\hline
KS-test (D) for & Significance for & KS-test (D) for & Significance for \\
z distribution & z distribution & peak flux distribution & peak flux distribution \\
\hline
$4.77 \times 10^{-2}$ & 99.79\% & $3.09^{+4.12}_{-1.04} \times 10^{-2}$ & $85.59^{+14.10}_{-81.93}\%$ \\
\hline\hline
\end{tabular}
\end{center}

\end{table}

\begin{figure}[!h]
\begin{center}
\includegraphics[width=0.84\textwidth]{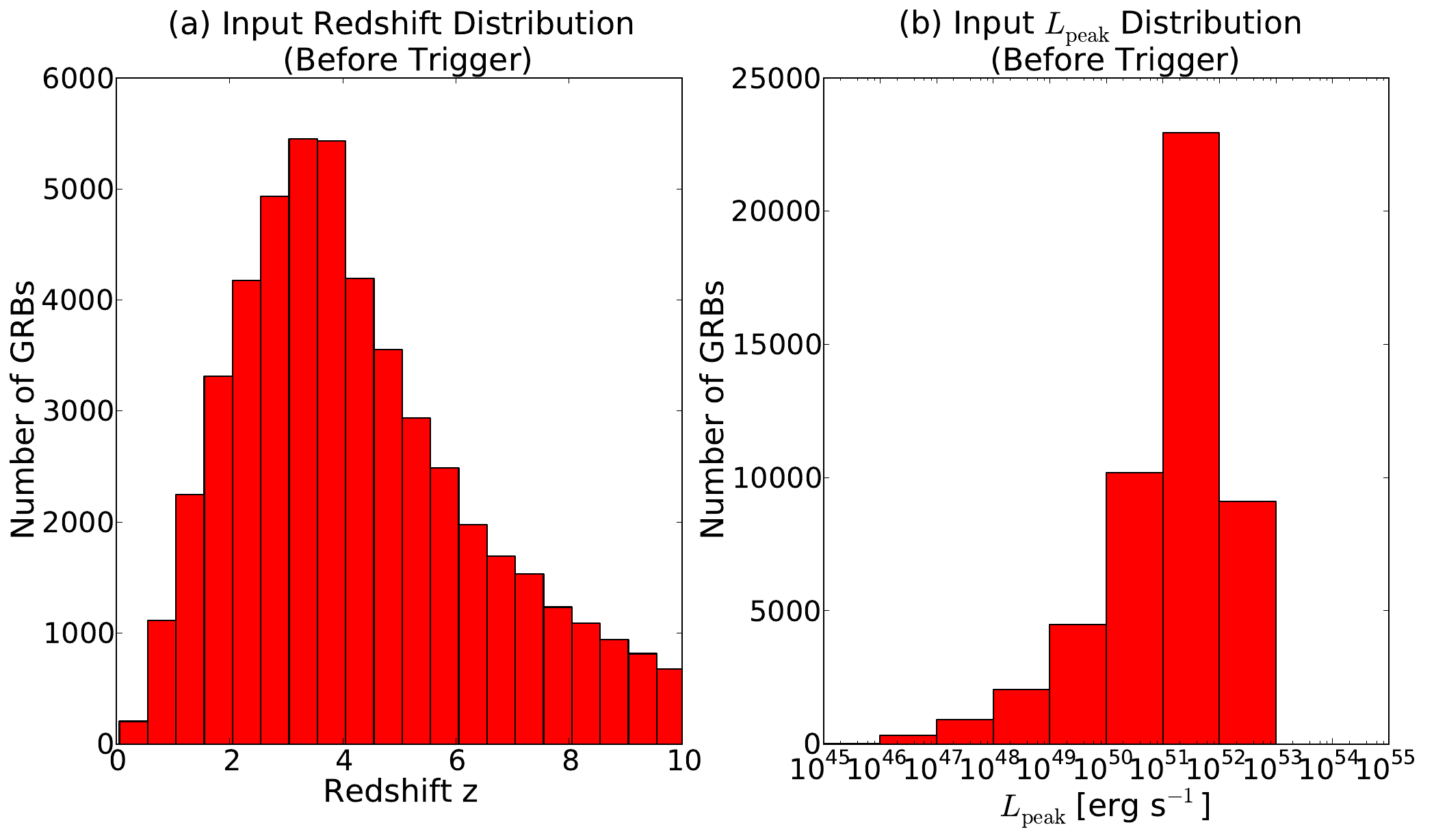}
\includegraphics[width=0.64\textwidth]{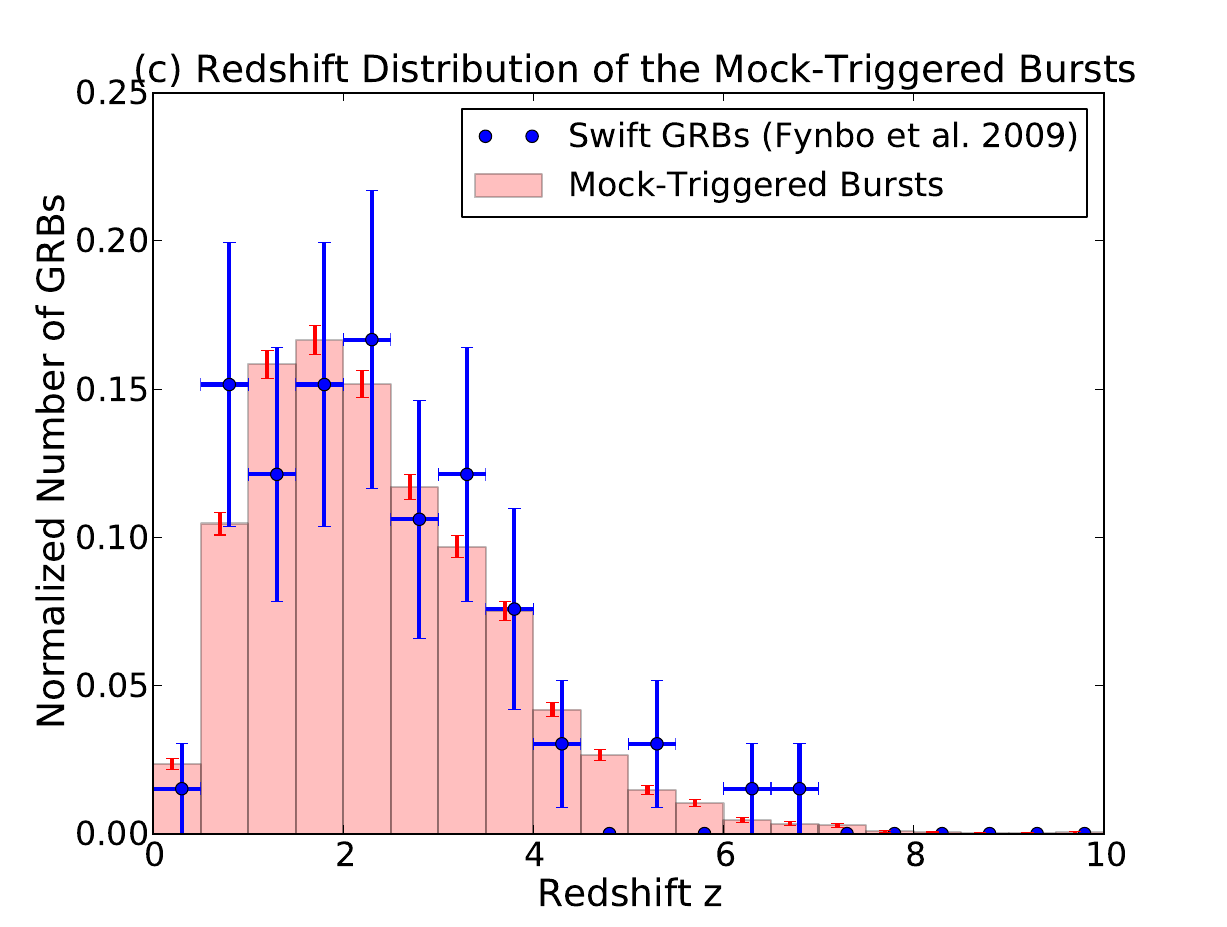}
\end{center}
\caption{
{\it Panel (a)}: the redshift distribution of all 50000 simulated bursts.
{\it Panel (b)}: the peak-luminosity distribution of all 50000 simulated bursts.
{\it Panel (c)}: the redshift distribution for the mock-triggered bursts (red bars), which are those bursts that are triggered by our trigger simulator.
The redshift distribution of the real BAT-detected bursts is also plotted for comparison (blue dots; \citet{Fynbo09}).
Error bars along the $y$-axis show the statistical errors in each bin. Error bars along the $x$-axis represent the bin sizes. 
}
\label{fig:redshift}
\end{figure}

Figure~\ref{fig:redshift} shows the redshift distribution from this best-fit sample.
Panels (a) and (b) of Fig.~\ref{fig:redshift} give the input redshift and luminosity distributions, respectively,
of all 50000 bursts created. 
Panel (c) of the figure shows the comparison of the redshift distribution between the mock-triggered bursts 
and real observations.
The red bars in panel (c) show the normalized numbers of
the simulated bursts that are triggered by our trigger-simulator.
The blue dots in panel (c) show the normalized numbers of real BAT detections. 
The error bars along the $y$-axis are the statistical errors (i.e., square root of the number in each bin).
The $x$-axis error bars simply represent the bin size.
As mentioned in Sect.~\ref{sect:BAT_redshift}, we adopt the GRB redshift sample 
reported in \citet{Fynbo09} for our comparison.

\begin{figure}[!h]
\begin{center}
\includegraphics[width=0.67\textwidth]{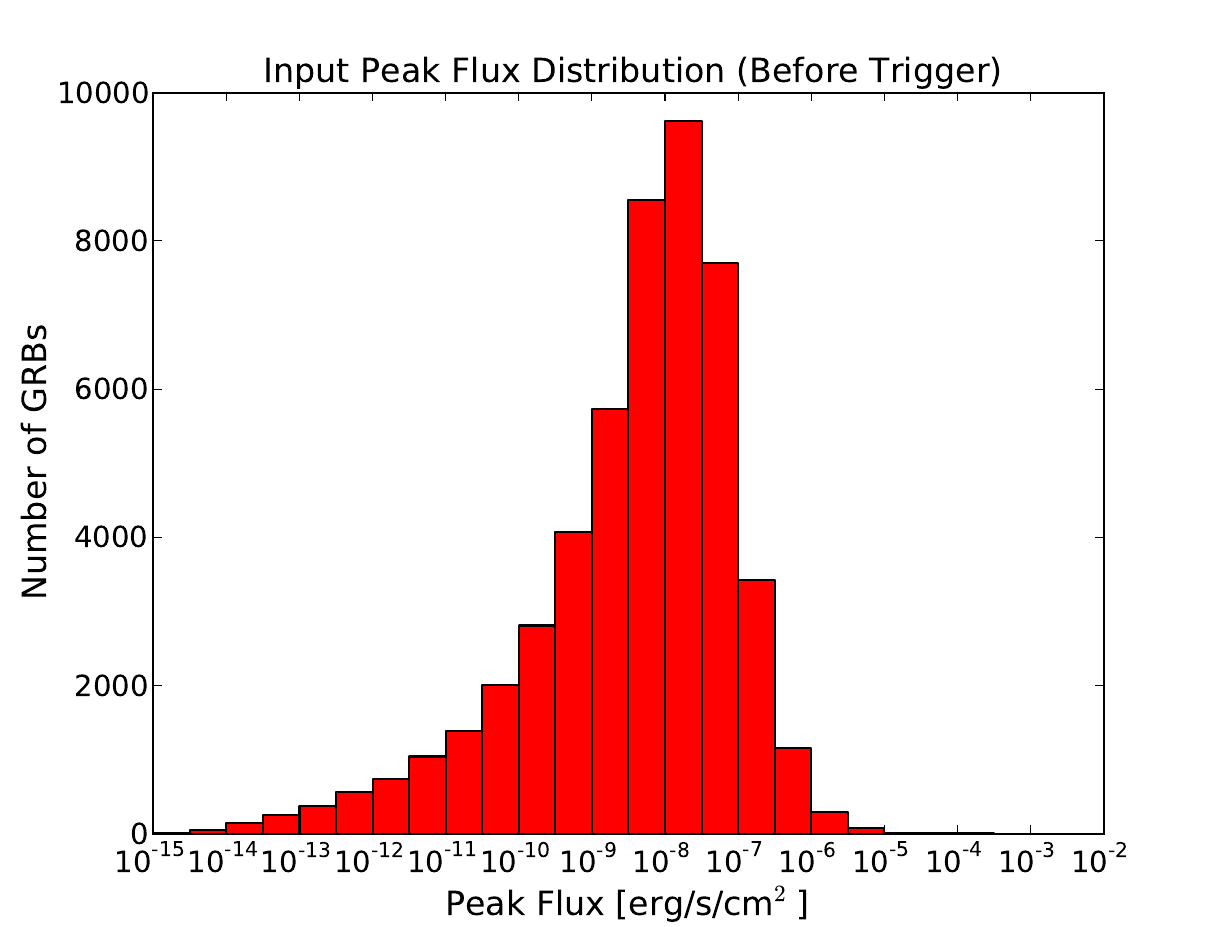}
\includegraphics[width=0.67\textwidth]{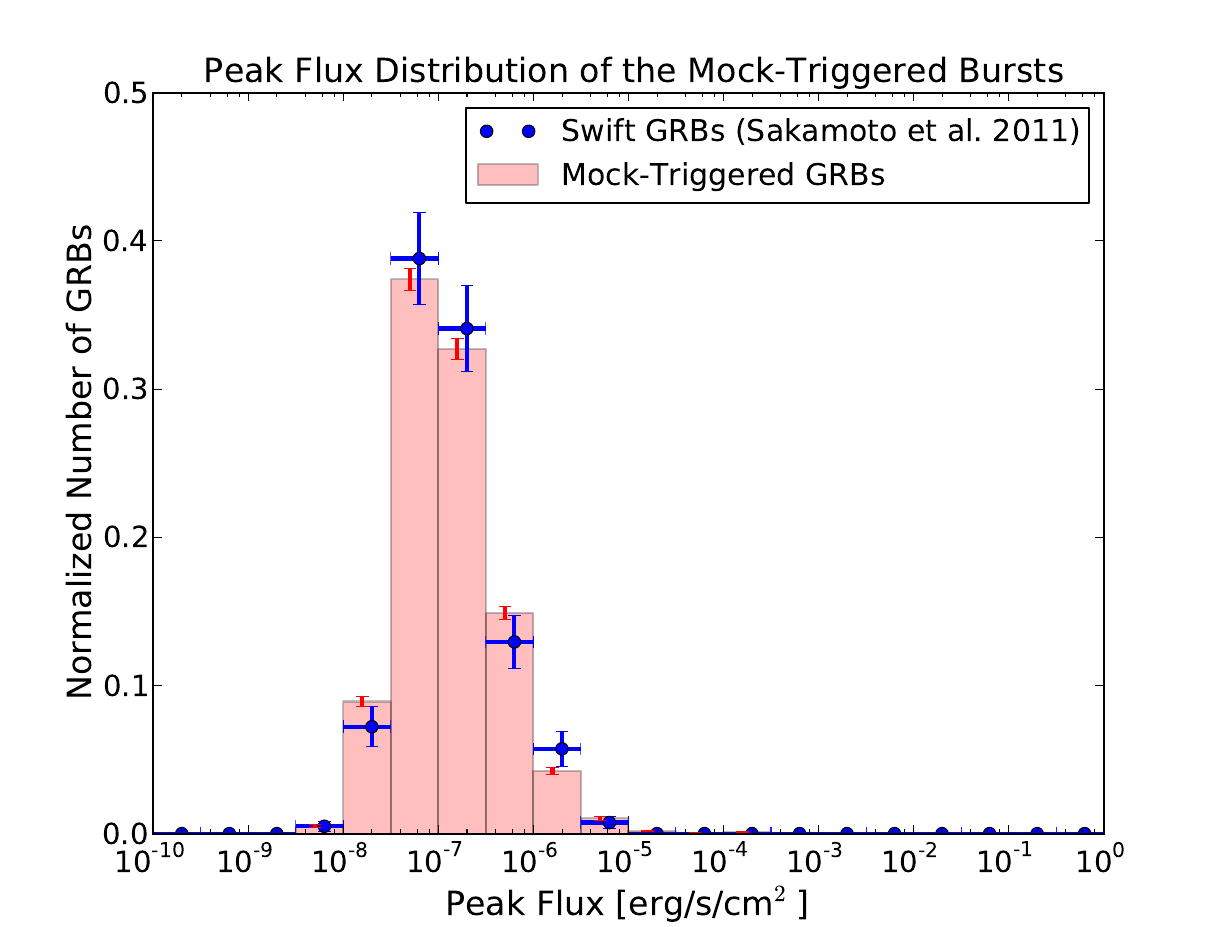}
\end{center}
\caption{
{\it Upper Panel}: the peak-flux distribution of all the 50000 simulated bursts.
{\it Bottom Panel}: the peak-flux distribution for the mock-triggered bursts (red bars).
The peak-flux distribution of the real BAT-detected bursts are also plotted for comparison (blue dots; \citet{Sakamoto11}).
Error bars along the $y$-axis show the statistical errors in each bin. Error bars along the $x$-axis represent the bin sizes.
}
\label{fig:peakflux}
\end{figure}

Figure~\ref{fig:peakflux} shows the peak-flux distribution from the best-fit sample.
The top panel gives the input peak-flux distribution from the whole sample of 50000 bursts.
The bottom panel shows the peak-flux distribution of the simulated bursts that 
are triggered by our trigger-simulator. Again, this mock-triggered sample is given
as red bars and the distribution of the real BAT-detected GRBs is shown as blue dots.
The peak fluxes of the real BAT-detected GRBs are reported in \citet{Sakamoto11}.
These authors also listed the errors for each measured peak flux. Therefore, 
we run a quick Monte Carlo simulation to see how the match of the two distributions 
(i.e., the value from the KS test)
changes if one allows the peak fluxes to change within their error bars.
Results show that the uncertainty in the peak-flux measurements
can change the ``D'' value in the KS test from
 $7.21 \times 10^{-2}$ to $2.05 \times 10^{-2}$, which correspond to 
 a significance level of 3.66\% and 99.69\%, respectively. 
 These are the uncertainties we listed in Table~\ref{tab:best_result}.

\begin{figure}[!h]
\begin{center}
\includegraphics[width=0.55\textwidth]{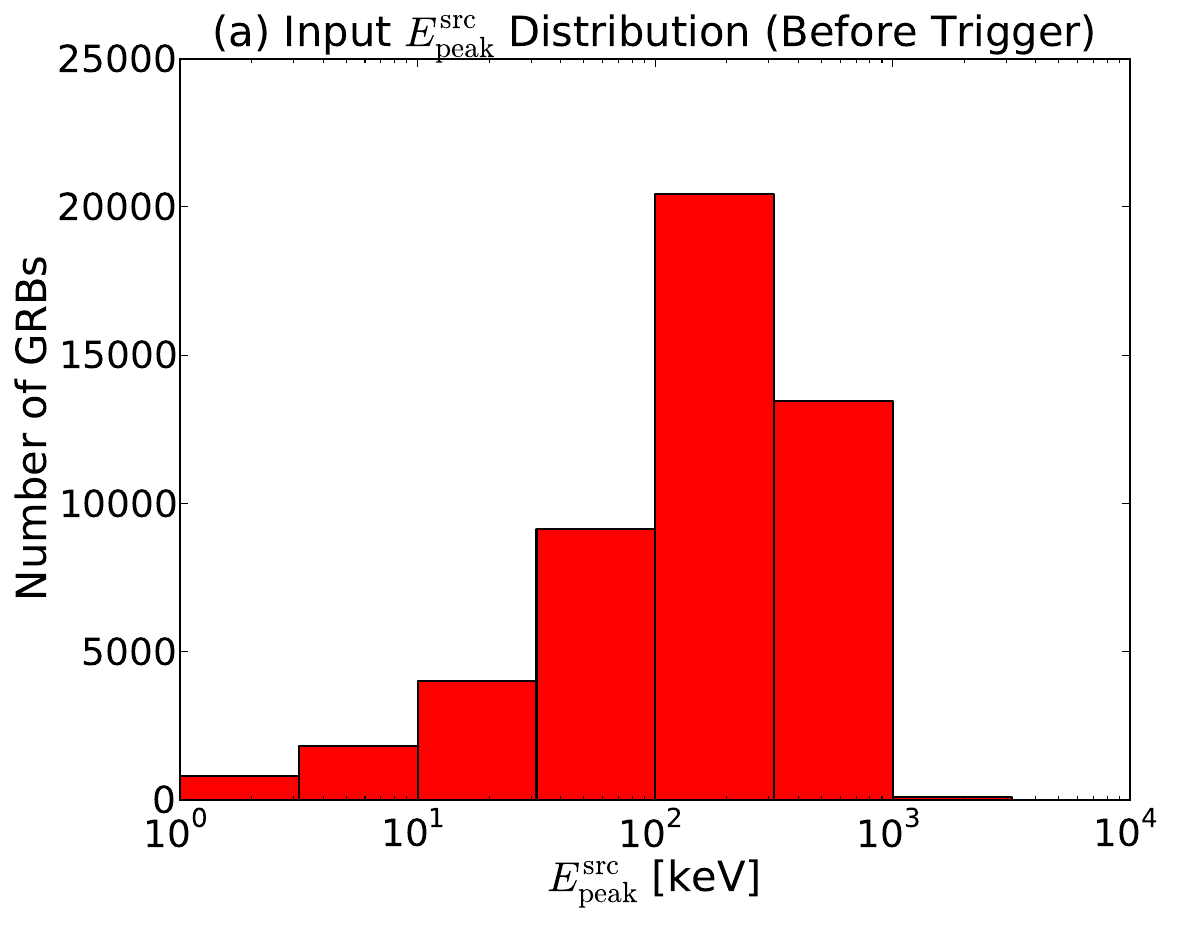}
\includegraphics[width=0.95\textwidth]{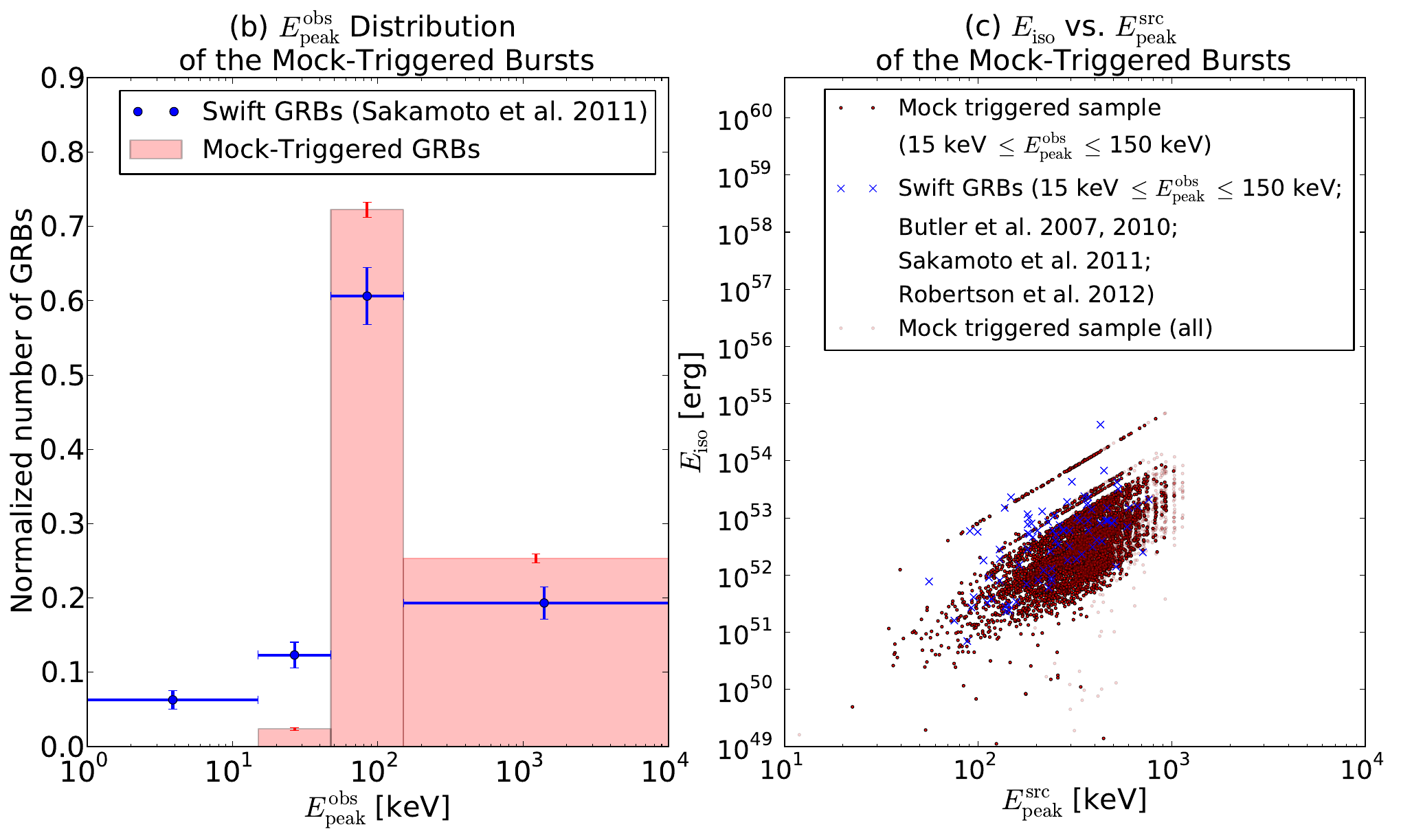}
\end{center}
\vspace{-30pt}
\caption{
{\it Panel (a)}: the $E^{\rm src}_{\rm peak}$ distribution of all 50000 simulated bursts.
{\it Panel (b)}: the $E^{\rm obs}_{\rm peak}$ distribution for the mock-triggered bursts (red bars).
The $E^{\rm obs}_{\rm peak}$ distribution of the real BAT-detected bursts are plotted (blue bars; \citet{Sakamoto11}).
Error bars along the $y$-axis show the statistical errors in each bin. Error bars along the $x$-axis represent the bin sizes.
Small and large bins are bursts wth $E^{\rm obs}_{\rm peak}$ inside and outside of the BAT energy range, respectively.
{\it Panel (c)}: $E_{\rm iso}$ versus $E^{\rm src}_{\rm peak}$.
Dark red dots and blue crosses show bursts with $E^{\rm obs}_{\rm peak}$ in the BAT energy range for the mock-triggered sample and the real GRBs, respectively. 
The full mock-triggered sample is shown as light red dots.
$E_{\rm iso}$ values in the plot are integrated over $T_{90}$.
}
\vspace{-40pt}
\label{fig:Epeak}
\end{figure}

Figure~\ref{fig:Epeak} plots the $E_{\rm peak}$ distribution for this best-fit sample and
its comparison with the distribution from real BAT-detected GRBs.
Panel (a) plots the input  $E^{\rm src}_{\rm peak}$ distribution of the 50000 simulated bursts.
Panel (b) shows the comparison of the $E^{\rm obs}_{\rm peak}$ distribution
between the mock-triggered bursts and the real BAT-detected GRBs.
Similarly, this mock-triggered sample is plotted
in red bars and the distribution of the real BAT-detected GRBs is shown as blue dots.
The $E^{\rm obs}_{\rm peak}$ values of the real GRBs are estimated from \citet{Sakamoto09, Sakamoto11},
as described in Sect.~\ref{sect:BAT_Epeak}.
Because it is hard to estimate $E^{\rm obs}_{\rm peak}$ when the values fall out 
of the BAT energy range ($\sim 15 - 150$ keV), we make two large bins for bursts with $E^{\rm obs}_{\rm peak}$
outside of the BAT range.
Inside the BAT energy range, the data are binned into two smaller bins.

\vspace{-10pt}

Panel(c) of Fig.~\ref{fig:Epeak} shows the $E_{\rm iso}$-$E^{\rm src}_{\rm peak}$ correlation.
Red dots in this plot represent values from the mock-triggered sample.
Dark red dots show the bursts with $E^{\rm obs}_{\rm peak}$ values inside the BAT energy range,
while light red dots are events with $E^{\rm obs}_{\rm peak}$ values outside of the BAT energy range.
Blue crosses show the bursts from the real BAT detections.
Due to the the large uncertainties of the $E^{\rm src}_{\rm peak}$ and $E_{\rm iso}$ values
for real bursts, only events with the observed $E^{\rm obs}_{\rm peak}$ inside the BAT energy range
are shown in the plot (see detailed discussion in Sect.~\ref{sect:BAT_observables}).
The $E_{\rm iso}$ values in the figure are integrated over $T_{90}$,
both for the real GRBs and the simulated bursts.
The intrinsic $E_{\rm iso}$ values of the real BAT-detected bursts 
are unknown due to background noise.
Therefore, we also restrict the $E_{\rm iso}$ to the $T_{90}$ range for 
the simulated bursts to have a fair comparison with real observations.



Results from our search suggest that 
a slightly modified Yonetoku relation
\begin{align}
\label{eq:Yonetoku_mod}
E^{\rm src}_{\rm peak} = 1.8 \times \left(\frac{1}{2.34 \times 10^{-5}} \times \frac{L_{\rm peak}}{10^{52} \ \rm erg \ s^{-1}}  \right)^{0.5}.
\end{align}
creates a better match for the $E_{\rm peak}$ distribution, as shown in Panel (b) of Fig.~\ref{fig:Epeak}.
Everything inside the bracket of Eq.~\ref{eq:Yonetoku_mod} is the original functional from \citet{Yonetoku04},
but we use peak luminosity $L_{\rm peak}$ instead of the isotropic luminosity $L_{\rm iso}$.
In our modified Yonetoku relation, $E^{\rm src}_{\rm peak}$ is 1.8 times larger than that produced by
the original Yonetoku relation.
The main reason for us to consider this modified Yonetoku relation is because this relation generates higher $E^{\rm src}_{\rm peak}$ 
bursts for the same $L_{\rm peak}$, and thus increases the number of detections of higher $E^{\rm src}_{\rm peak}$
bursts and matches the observations better.

This best-fit sample predicts that {\it Swift} should detect $\sim 96$ bursts per year,
which is in good agreement with the average number of $\sim 95$ GRBs per year from 2005 to 2009 based on real BAT observations \citep{Sakamoto11}.
The predicted detection rate for {\it Swift} ($R_{Swift}$)
is calculated based on the following equation,
\begin{align}
R_{Swift} = {\cal R}_{\rm GRB; dz} \times f_{\rm detect} \times \rm FOV \times \rm t_{\rm survey},
\end{align}
where ${\cal R}_{\rm GRB; dz}$ is the observed GRB rate in units of number per redshift per solid angle per time in the observed frame
(Eq~\ref{eq:GRBrate_dz}),  
$f_{\rm detect}$ is the detection rate, as shown in the third column of Table~\ref{tab:epeak_noevo},
FOV $\sim 2 \ \rm sr$ is the field-of-view of the BAT \citep{Barthelmy05},
and $t_{\rm survey} \sim 90\%$ is the fraction of the time that BAT spends on searching for GRBs.

As discussed in Sect.~\ref{sect:general_methods}, we rerun this best-fit sample with the complete set of long trigger criteria,
including those with foreground periods shorter than two seconds,
in order to make sure the results remain a good fit with observations.
We use 26884 active detectors in this run, which is the average number of active detectors from 2005 to 2009. 
Results show that comparing with the same sample (with 26884 active detectors) using only the trigger criteria longer than two seconds,
adopting the complete set of trigger criteria, change the detection rate from 13.73\% to 14.01\%.
The KS test significance for the redshift distribution changes from 99.02\% to 99.32\%,
and the KS test significance for the flux distribution changes from $81.80^{+17.91}_{-65.55}\%$
to $89.78^{+9.41}_{-67.80}\%$.
All of these changes are significantly smaller than the statistical uncertainty. 
Therefore, results using our best-fit parameters remain good matches with observations
when adopting the full set of long trigger criteria.

\section{$E^{\rm src}_{\rm peak}$ distribution and evolution}
\label{sect:Epeak}

\subsection{$E^{\rm src}_{\rm peak}$ distribution}
\label{sect:Epeak_dist}

The intrinsic $E^{\rm src}_{\rm peak}$ distribution remains uncertain and controversial. Many studies have 
suggested possible correlations between the total energy output of the burst and the peak energy of the $\nu F_{\nu}$ spectrum in the burst rest frame $E^{\rm src}_{\rm peak}$.
These relations
often attempt to relate $E^{\rm src}_{\rm peak}$ and $E_{\rm iso}$, the total energy fluence of the burst,
or $E^{\rm src}_{\rm peak}$ and $L_{\rm iso}$, the total luminosity of the burst \citep{Amati02, Amati06, Ghirlanda04,Yonetoku04}. 

In order to see the consequences of adopting different $E^{\rm src}_{\rm peak}$ distributions, 
we test several functions that have significantly different shapes compare to those used in our best-fit sample (described in Sect.~\ref{sect:rate}).
The distributions we test include:
(1) A flat $E^{\rm src}_{\rm peak}$ distribution in linear space.
(2) A flat $E^{\rm src}_{\rm peak}$ distribution in logarithmic space. 
(3) A Gaussian $E^{\rm src}_{\rm peak}$ distribution in logarithmic space, with average $= 300$ keV and $\sigma = 1$.
(4) A special function 
\begin{align}
\phi (\rm log_{10}(\it E^{\rm src}_{\rm peak})) \propto \left \{ \begin{array}{ll}
						\rm{log}_{10}(\it E^{\rm src}_{\rm peak}) & \mbox{if} \ E^{\rm src}_{\rm peak} < 10 \ \mbox{keV}, \\
						\rm{log}_{10}(10 \ \rm keV) & \mbox{otherwise.}
						\end{array} \right.
\end{align}				
This function contains a lower number of bursts for $E^{\rm src}_{\rm peak} < 10$ keV 
and follows a flat distribution in logarithmic space for events with $E^{\rm src}_{\rm peak} > 10$ keV. 
$E^{\rm obs}_{\rm peak}$ evolution is not included in these simulations.

Table~\ref{tab:epeak_noevo} summarizes the major results from using different $E^{\rm src}_{\rm peak}$ distributions.
All the results, except our best-fit sample (the one with a modified Yonetoku $E^{\rm src}_{\rm peak}$ distribution), 
are based on 10000 simulated bursts and using 26884 active detectors, which is the average number of 
active detectors from year 2005 to 2009 (the time period of the real BAT-detected bursts adopted in this paper). 
Our best-fit result contains a total number of 50000 simulated bursts with different numbers of enabled detectors, as described in Sect.~\ref{sect:best_result}.

\begin{figure}[!h]
\begin{center}
\includegraphics[width=1.0\textwidth]{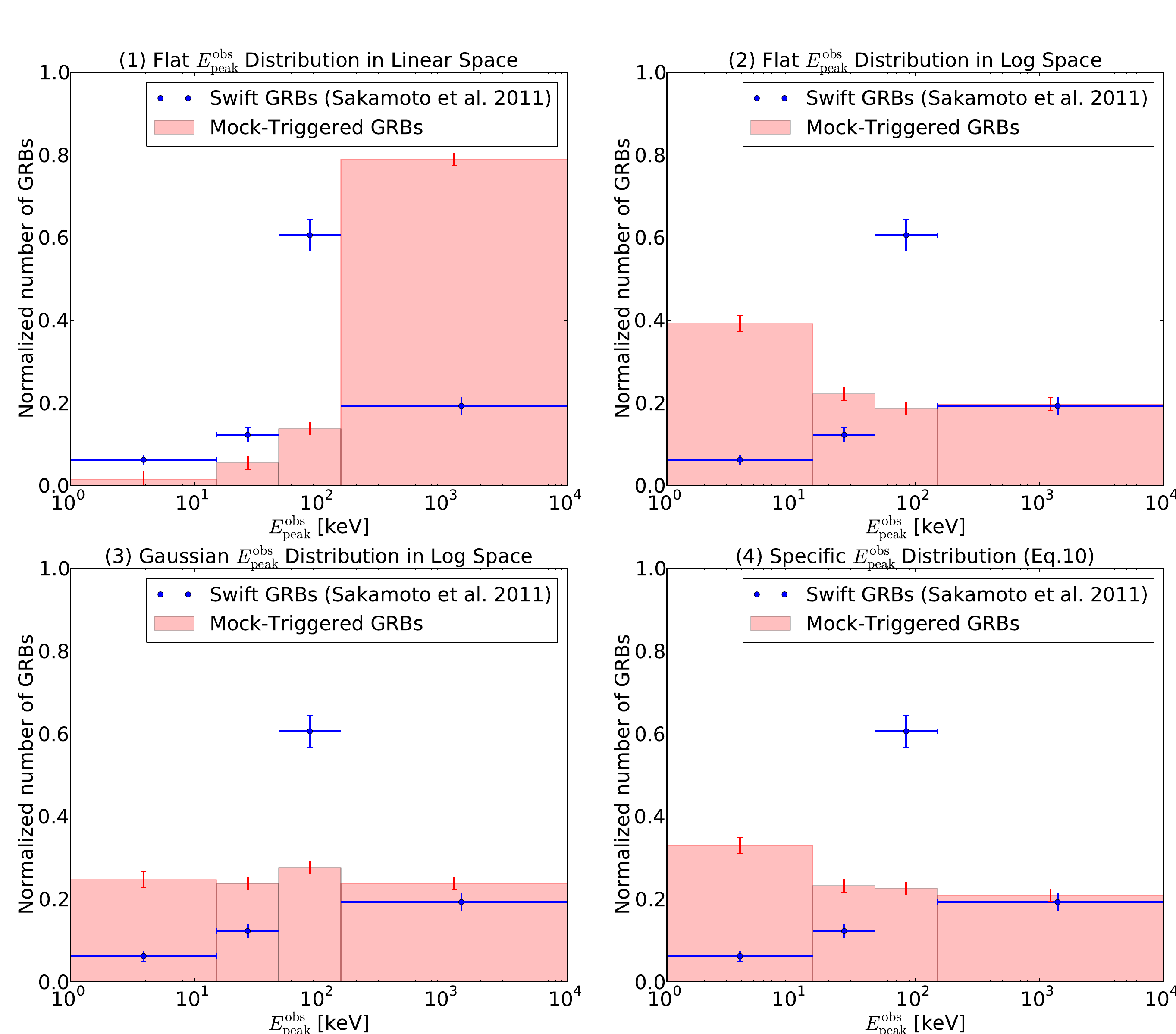}
\end{center}
\vspace{-20pt}
\caption{
Comparison of the $E^{\rm obs}_{\rm peak}$ distributions for the mock-triggered bursts (red bars)
assuming different input $E^{\rm src}_{\rm peak}$ distributions (Sect.~\ref{sect:Epeak_dist}).
The $E^{\rm obs}_{\rm peak}$ distribution of the real BAT-detected bursts are also plotted for comparison (blue bars; \citet{Sakamoto11}).
The narrow bins indicate bursts with $E^{\rm obs}_{\rm peak}$ values inside the BAT detectable energy range.
The two large bins contain bursts with $E^{\rm obs}_{\rm peak}$ values outside of the BAT energy range. 
Error bars along the $y$-axis show the statistical errors in each bin. Error bars along the $x$-axis represent the bin sizes. }
\label{fig:logEpeak_allexample}
\vspace{-10pt}
\end{figure}

\begin{figure}[!h]
\begin{center}
\includegraphics[width=1.0\textwidth]{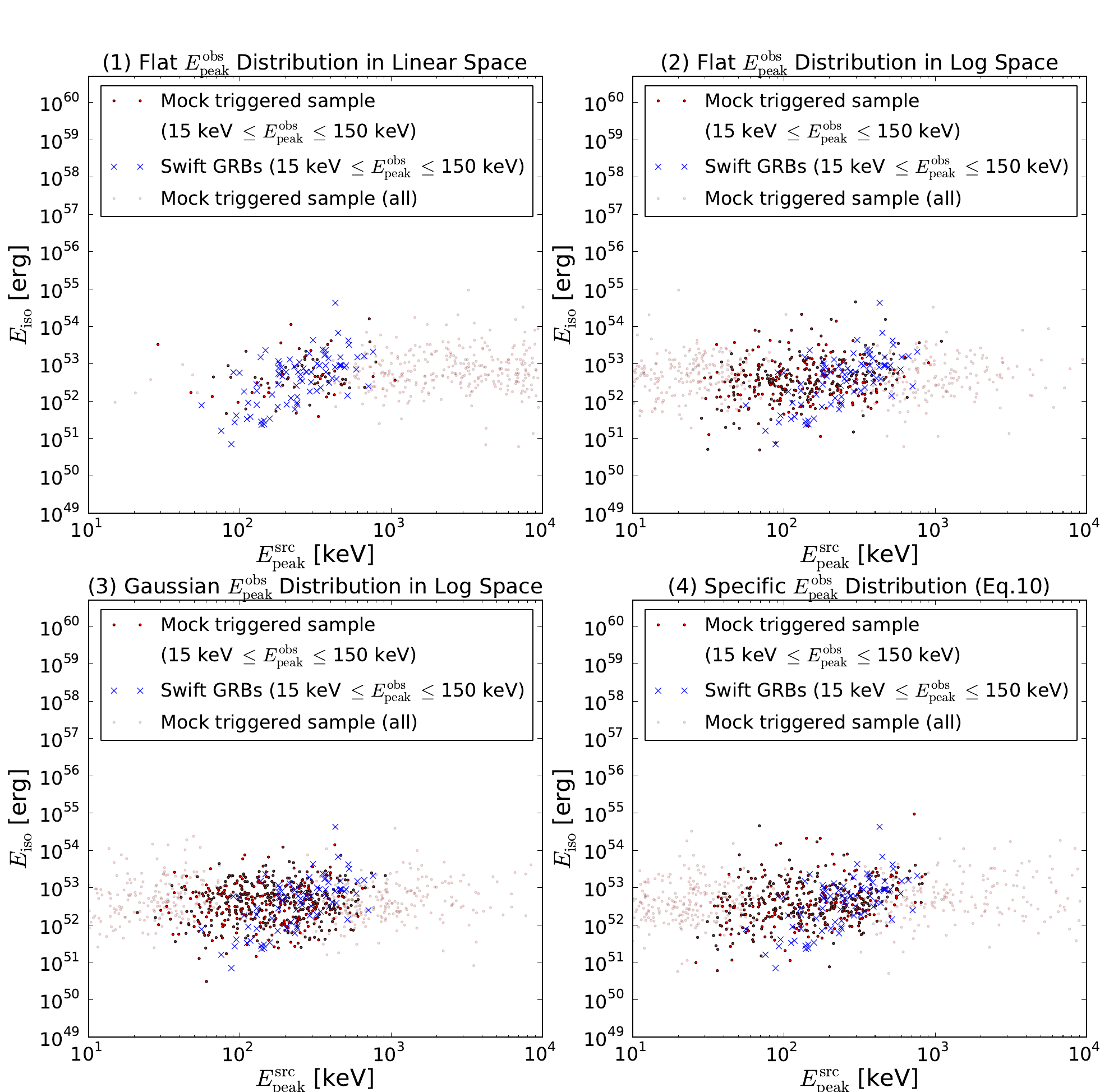}
\end{center}
\caption{
Comparison of the resulting $E_{\rm iso}$ versus $E^{\rm src}_{\rm peak}$ for the mock-triggered bursts
using different input $E^{\rm src}_{\rm peak}$ distributions (Sect.~\ref{sect:Epeak_dist}).
Dark red dots and blue crosses represent bursts with $E^{\rm obs}_{\rm peak}$ in the BAT energy range for the mock-triggered sample 
and the real BAT-detected GRBs, respectively. 
$E^{\rm obs}_{\rm peak}$ for real GRBs are from \citet{Butler07, Butler10, Sakamoto11}, and \cite{Robertson12}.
For comparison, the full mock-triggered sample is also shown as light red dots.
$E_{\rm iso}$ in the plot represents $E_{\rm iso}$ in the $T_{90}$ range.}
\label{fig:Epeak_Eiso_allexample}
\end{figure}

For all the four different $E^{\rm src}_{\rm peak}$ distributions we test, 
the resulting redshift and peak flux distributions of the mock-triggered samples remain good fits to the real observations.
In fact, for the peak-flux distributions, for which we can quantify the uncertainties of the fit, 
the KS-tests show that all these matches based on different $E^{\rm src}_{\rm peak}$ distributions are within the calculated uncertainties of each other.
In other words, all the $E^{\rm src}_{\rm peak}$ distributions in our tests give the same level of good fits to the observations,
and thus comparing the results to the redshift and peak-flux distributions alone would not be sufficient to distinguish
different $E^{\rm src}_{\rm peak}$ distributions.
However, using different $E^{\rm src}_{\rm peak}$ distributions does change the detection rates.
For the four different cases we tried, the detection rate can change from $\sim 4\%$ to $\sim 14\%$,
and thus will affect the normalization of the GRB rate (i.e., the GRB rate at $z=0$) up to $\sim$ 4 times higher.

The major distinctions of adopting different $E^{\rm src}_{\rm peak}$ distributions appear in the $E_{\rm peak}$ distributions 
of the mock-triggered samples. 
Figure~\ref{fig:logEpeak_allexample} and \ref{fig:Epeak_Eiso_allexample} 
compare the resulting $E^{\rm obs}_{\rm peak}$ distributions and 
the $E_{\rm iso}-E^{\rm src}_{\rm peak}$ correlations based on these four different $E^{\rm src}_{\rm peak}$ distributions.
As one can see from these figures, 
all these additional
$E^{\rm src}_{\rm peak}$ distributions we test seem to generate worse matches to the observational sample 
than the modified Yonetoku sample (see Fig.~\ref{fig:Epeak}),
in both the $E^{\rm obs}_{\rm peak}$ distributions and the $E_{\rm iso}$-$E^{\rm src}_{\rm peak}$ correlations.
However, due to the large and hard-to-quantify uncertainties in the $E^{\rm obs}_{\rm peak}$ and $E_{\rm iso}$ of the real observational sample,
it remains ambiguous whether any of these $E^{\rm src}_{\rm peak}$ distributions can be excluded.
Therefore, these plots are only shown to indicate how the distributions can change if one assumes different $E^{\rm src}_{\rm peak}$ distributions;
no conclusion about the intrinsic $E^{\rm src}_{\rm peak}$ distribution can be drawn until 
direct measurements of $E^{\rm obs}_{\rm peak}$ and $E_{\rm iso}$
become available.

\begin{table}[ht]
\caption{\label{tab:epeak_noevo}
Summary of the results from simulations without $E^{\rm obs}_{\rm peak}$ evolution. The five samples shown here are 
based on different $E^{\rm src}_{\rm peak}$, as described in Sect.~\ref{sect:Epeak_dist}.
\vspace{-10pt}
}
\begin{center}
\begin{tabular}{|c|c|c|c|c|}
\hline\hline
$E^{\rm src}_{\rm peak}$ & Detection & Prediction for          & KS-test significance & KS-test significance for \\
Distribution                         & Rate          & {\it Swift} [$\rm yr^{-1}$] & for z distribution & peak flux distribution \\\hline
Modified Yonetoku & 13.95\% & 95.60 & 99.79\% & $85.59^{+14.10}_{-81.93}\%$ \\
Flat in Linear Space & 3.76\% & 25.76 & 14.88\% & $88.81^{+10.35}_{-69.58}\%$ \\  
Flat in Log Space & 8.54\% & 58.51 & 68.24\% & $39.95^{+59.41}_{-34.21}\%$ \\
Gaussian & 9.74\% & 66.73 & 79.08\% & $79.34^{+20.65}_{-76.50}\%$ \\
Specified Function & 9.05\% & 62.00 & 63.18\% & $96.32^{+3.68}_{-83.07}\%$ \\
\hline\hline
\end{tabular}
\end{center}
\end{table}
\begin{table}[ht]
\caption{\label{tab:epeak_evo}
Summary of the results from simulations with $E^{\rm obs}_{\rm peak}$ evolution. The five samples shown here are 
based on different $E^{\rm src}_{\rm peak}$, as described in Sect.~\ref{sect:Epeak_dist}.
\vspace{-10pt}
}
\begin{center}
\begin{tabular}{|c|c|c|c|c|}
\hline\hline
$E^{\rm src}_{\rm peak}$ & Detection & Prediction for          & KS-test significance & KS-test significance for \\
Distribution                         & Rate          & {\it Swift} [$\rm yr^{-1}$] & for z distribution & peak flux distribution \\
\hline
Modified Yonetoku & 15.05\% & 103.10 & 98.62\% & $69.34^{+20.00}_{-62.60}\%$ \\
Flat in Linear Space & 4.65\% & 31.86 & 31.39\%& $0.32^{+2.02}_{-0.32}\%$ \\
Flat in Log Space & 8.63\% & 59.12 & 59.15\% & $47.88^{+51.11}_{-41.73}\%$ \\
Gaussian &10.10\% & 69.19 & 76.84\% & $50.68^{+48.47}_{-49.47}\%$ \\
Specified Function & 9.37\% & 64.19 & 57.44\% & $89.00^{+10.95}_{-80.81}\%$ \\
\hline\hline
\end{tabular}
\end{center}
\end{table}

\subsection{$E^{\rm obs}_{\rm peak}$ evolution}
\label{sect:Epeak_evo}

As discussed in Sect.~\ref{sect:obs_lc}, spectral evolution has been observed in many GRBs,
and thus we implement an option to include $E^{\rm obs}_{\rm peak}$ evolution in our simulation.
To see how the results might change if $E^{\rm obs}_{\rm peak}$ evolution is included,
we apply this option to the best-fit parameter set (Table~\ref{tab:best_result}) 
with all five different $E^{\rm src}_{\rm peak}$ distributions we test.

Results with $E^{\rm obs}_{\rm peak}$ evolution included are summarized in Table~\ref{tab:epeak_evo}.
Similar to those samples presented in Table~\ref{tab:epeak_noevo}, results are from simulations of 
10000 bursts and using the average number of 26884 active detectors from year 2005 to 2009. 
These simulations show that including $E^{\rm obs}_{\rm peak}$ evolution can result in noticeable changes 
of the outcome, especially in the distributions of the burst characteristics.
Although our best-fit sample (the one with a modified Yonetoku $E^{\rm src}_{\rm peak}$ distribution)
remains good matches with both the redshift and peak-flux distributions,
samples with other $E^{\rm src}_{\rm peak}$ distributions show that 
the resulting KS-test values can change considerably with $E^{\rm obs}_{\rm peak}$ evolution included.
However, we noted that when actually plotting these distributions, 
the changes do not seem to be as significant as indicated by the KS-test significance levels 
in the tables.
The general shapes and widths of the distributions remain similar with or without $E^{\rm obs}_{\rm peak}$ evolution.
This is because 
the KS-test values are especially sensitive to the medium of the distribution.
Therefore, a slight changes in the medium can lead to a major difference 
in the significance level.
Moreover, if the uncertainties in the significance levels of the peak-flux distributions are taken into account,
the data from real observations actually cannot distinguish the difference
between the results with or without $E^{\rm obs}_{\rm peak}$ evolution.

\begin{figure}[!h]
\begin{center}
\includegraphics[width=0.8\textwidth]{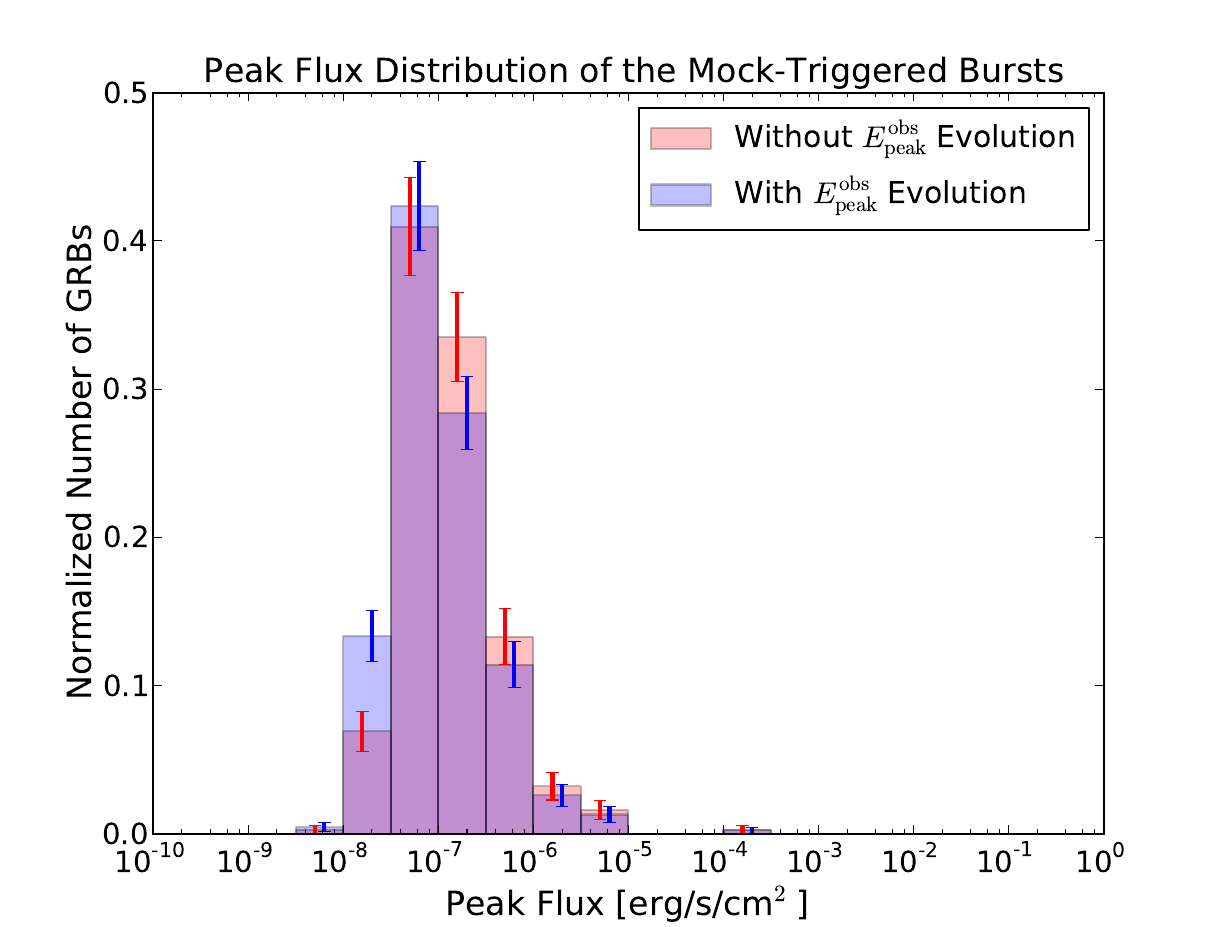}
\end{center}
\caption{
Comparison of the peak-flux distributions with and without $E^{\rm obs}_{\rm peak}$ evolution
for the sample using a flat $E^{\rm src}_{\rm peak}$ distribution in linear space.
The result with $E^{\rm obs}_{\rm peak}$ evolution is plotted as red bars.
The distribution without $E^{\rm obs}_{\rm peak}$ evolution is plotted as blue bars.
Error bars along the $y$-axis show the statistical errors in each bin. Error bars along the $x$-axis represent the bin sizes.
}
\label{fig:Epeak_evo}
\end{figure}

The sample using a flat $E^{\rm src}_{\rm peak}$ distribution in linear space
(see description in Sect.~\ref{sect:Epeak_dist})
shows the most remarkable change in the KS-test significance level when
 $E^{\rm obs}_{\rm peak}$ evolution is included. 
Therefore, Fig.~\ref{fig:Epeak_evo} plots the peak-flux distributions with and without $E^{\rm obs}_{\rm peak}$ evolution
of this sample as an example of how the general shapes of the distributions change.

When including the $E^{\rm obs}_{\rm peak}$ evolution, the detection rates in these samples either stay similar or 
increase slightly compared to those without $E^{\rm obs}_{\rm peak}$ evolution. 
A possible explanation is that as the burst spectra get softer, some of the bursts might have 
$E^{\rm obs}_{\rm peak}$ fall into the BAT energy range, and thus become detectable.
However, due to the low number of samples with $E^{\rm obs}_{\rm peak}$ evolution in our simulations,
we cannot exclude the possibility that this trend is a mere coincidence.

\section{Selection biases of using single trigger criterion}
\label{sect:biases}


To explore possible selection biases when using a single trigger criterion,
we use our best-fit sample (discussed in Sect.~\ref{sect:rate}) as an example 
and
calculate the detection fraction for each criterion.
The detection fraction here refers to the number of simulated bursts detected by each criterion divided 
by the total number of detections (when using all criteria).
In other words,
the detection fraction is the fraction of the mock-triggered bursts that would be detected if only using one criterion.
Because most of the bursts are detected by more than one criterion, the detection fractions from all criteria do not
add up to one.
Results show that the detection fraction for each criterion varies from $\sim 83\%$ for the most efficient criterion (Trigger Criterion \# 334)
to $\sim 19\%$ for the least efficient criterion (Trigger Criterion \# 356). 
Therefore, adopting the complex trigger algorithm of BAT increases the detection rate by 1.2 to 5.3 times.
The energy band $25-100$ keV is the most sensitive energy range of BAT,
which matches well with the peak of the BAT's effective area.

\begin{figure}[!h]
\begin{center}
\includegraphics[width=0.8\textwidth]{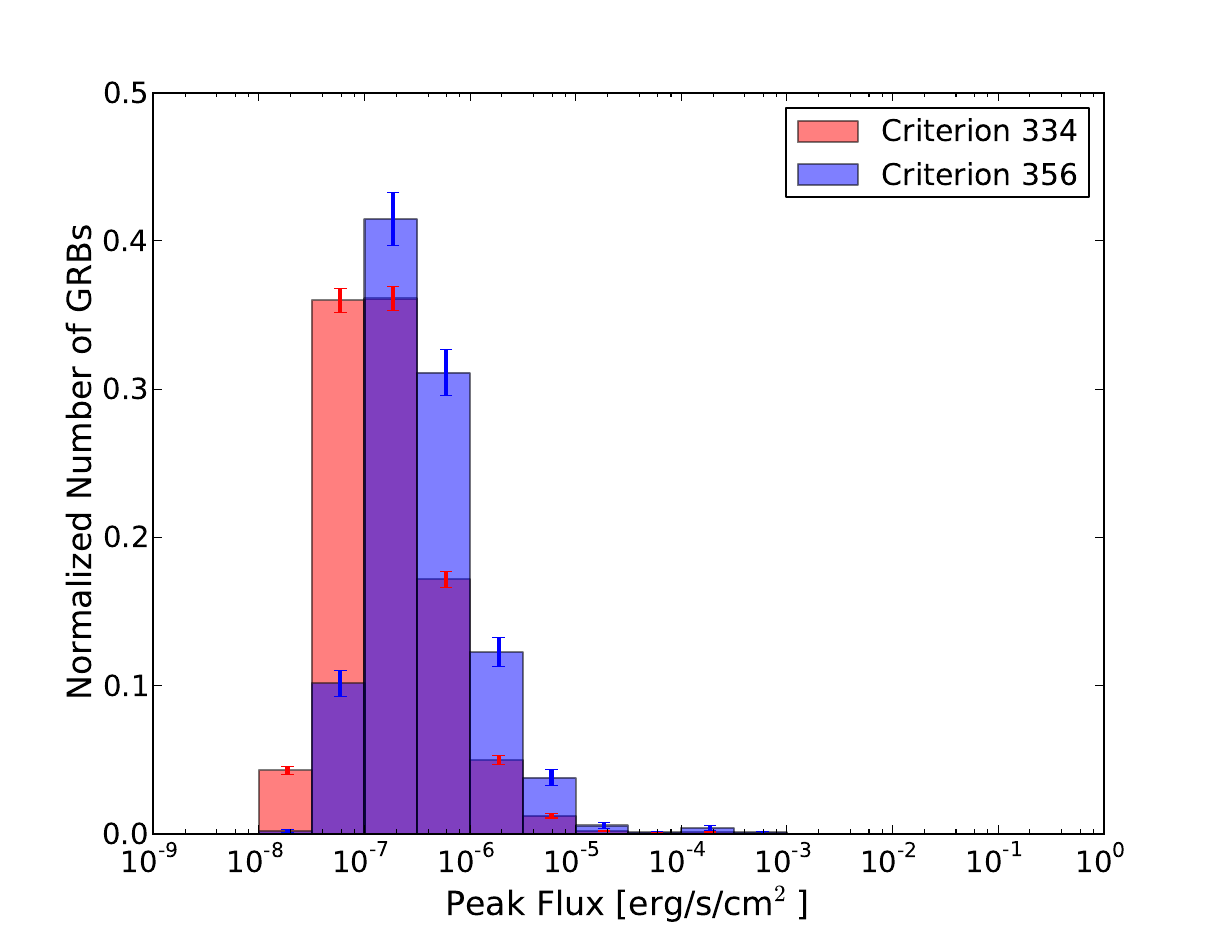}
\end{center}
\caption{
The peak-flux distributions of the simulated bursts triggered by the most efficient trigger criterion (Criterion 334) and the least efficient trigger criterion (Criterion 356).
Error bars along the $y$-axis show the statistical errors in each bin.
}
\label{fig:criterion_peakflux}
\end{figure}

Figure.~\ref{fig:criterion_peakflux} plots the peak-flux distributions of the simulated bursts triggered by the most
efficient criterion (Trigger Criterion \# 334) in red bars, and those triggered by the least efficient criterion (Trigger Criterion \# 356)
in blue bars. As one can see in the figure, the most efficient criterion is much more sensitive to the low-flux bursts 
than the least efficient criterion. 
Therefore, it is possible to miss some dim bursts when using 
only one single trigger criterion, and 
results in finding GRB characteristics that are biased toward the brighter end.
The complex trigger algorithm of BAT generates more detections of the low-flux events,
and thus improves the survey sensitivity.


\section{GRB detection fraction of the BAT-trigger algorithm}
\label{sect:detection_rate}

To generally understand how the complex BAT-trigger algorithm
affects burst detections,
we calculate the detection rate as a function of redshift.
From redshift $z = 0$ to $10$, we generate two hundred GRBs in each redshift bin
with bin size of $\Delta z = 0.2$.
This is to make sure each redshift bin has enough bursts to create a statistically meaningful result.
The redshift distribution of the bursts in each bin are uniformly assigned.
Other than the redshift, all other burst characteristics (such as luminosity function, spectral distribution, etc),
are the same as those found in our best-fit sample (Sect.~\ref{sect:best_result}).

\begin{figure}[!h]
\begin{center}
\includegraphics[width=0.8\textwidth]{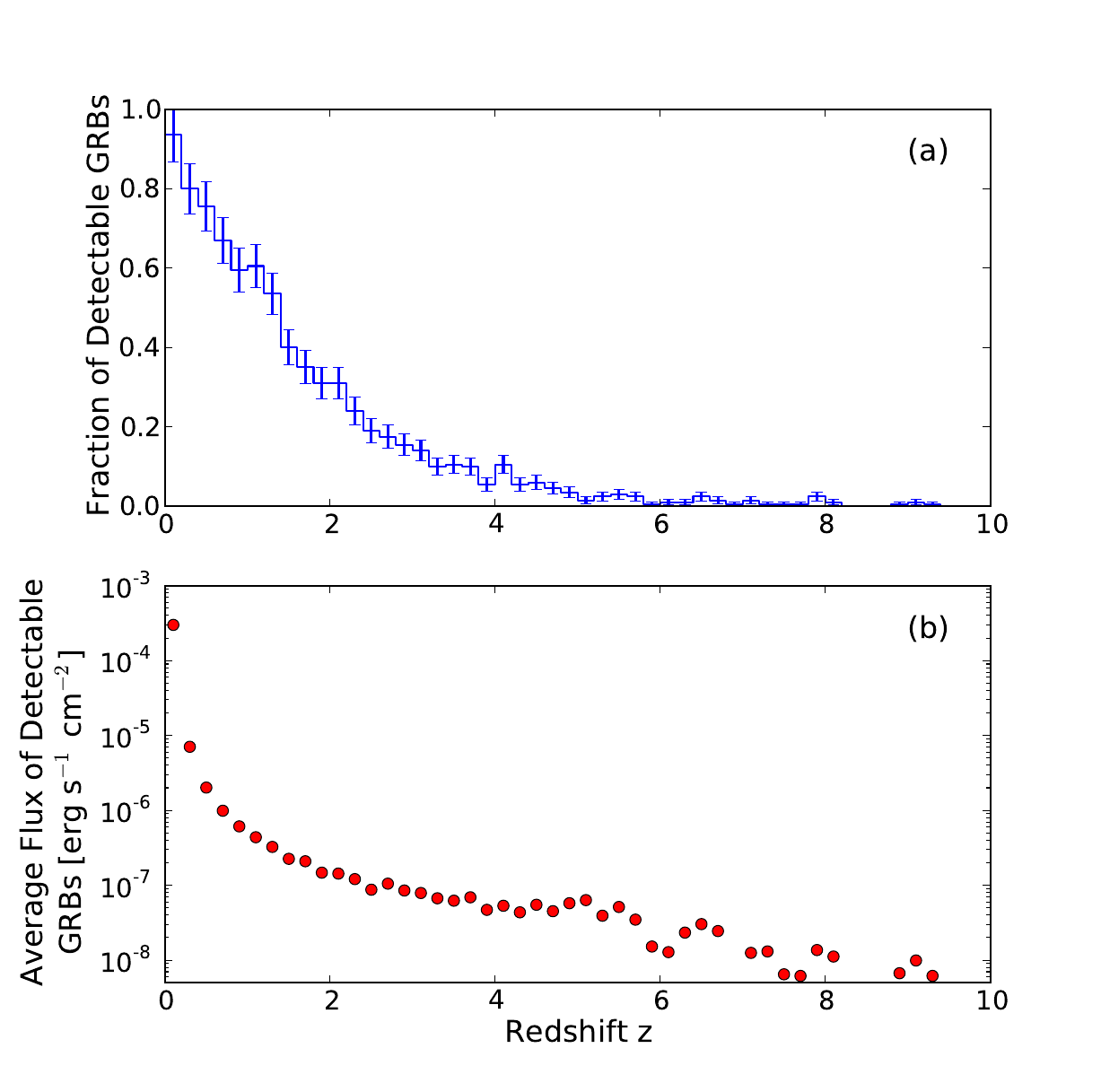}
\end{center}
\caption{
{\it Panel (a)}: The fraction of detectable GRBs as a function of redshift. 
The result is based on an intrinsically flat GRB redshift distribution, with other GRB characteristics (e.g., luminosity and spectral distributions) from our best-fit sample 
(Sect.~\ref{sect:best_result}). Error bars along the $y$-axis show the statistical errors in each bin.
{\it Panel (b)}: The average flux of the detectable GRBs in the simulation, as a function of redshift.
The adopted bin size is $\Delta z = 0.2$.
}
\label{fig:detection_rate}
\end{figure}

The results are plotted in Fig.~\ref{fig:detection_rate}.
Panel (a) shows the fraction of detectable bursts as a function of redshift.
Error bars along the $y$-axis show the statistical error in each bin (i.e., $\sqrt{N}$, where N is the number of bursts in each bin).
As expected, the detectable fraction is $\sim 1$ at $z \sim 0$, 
and drops to approximately zero at high redshift. 
Above redshift $z = 6$, the detectable fraction is about $1 - 3\%$.
Panel (b) presents the average flux of the detectable bursts in each redshift bin.
The average flux of the detectable bursts decreases 
with redshift as anticipated, since bursts become dimmer when they are further away. 
At redshift $z > 6$, the average flux of the detectable bursts
is around $10^{-8} \rm \ erg \ s^{-1} \ cm^{-2}$,
which is usually only detectable when the burst appears 
on-axis relative to the detector's plane.
There remain some statistical fluctuations in the plot, particularly at high redshift due
to the small number of detections.

\section{Comparison with the star-formation rate}
\label{sect:sfr}

As discussed in Sect.~\ref{sect:intro}, long GRBs are related to at least a sub-class of core-collapse supernovae.
Core-collapse supernovae are expected to directly trace the star-formation history due to the short lifetimes
of their massive progenitors. 
Thus, the cosmic long GRB rate is expected to follow the shape of the cosmic SFR.
\citet{Kistler08_SFR} noted that from the high-redshift GRB observations,
there is an unexpected rise in the cosmic GRB rate at large redshift 
compared to that expected from previous SFR measurements.
\citet{Yuksel08} use the high-redshift GRB measurements to calculate the high-redshift SFR, 
and conclude that SFR in the early universe might be larger than previously expected \citep{Hopkins06}.

\begin{figure}[!h]
\begin{center}
\includegraphics[width=1.0\textwidth]{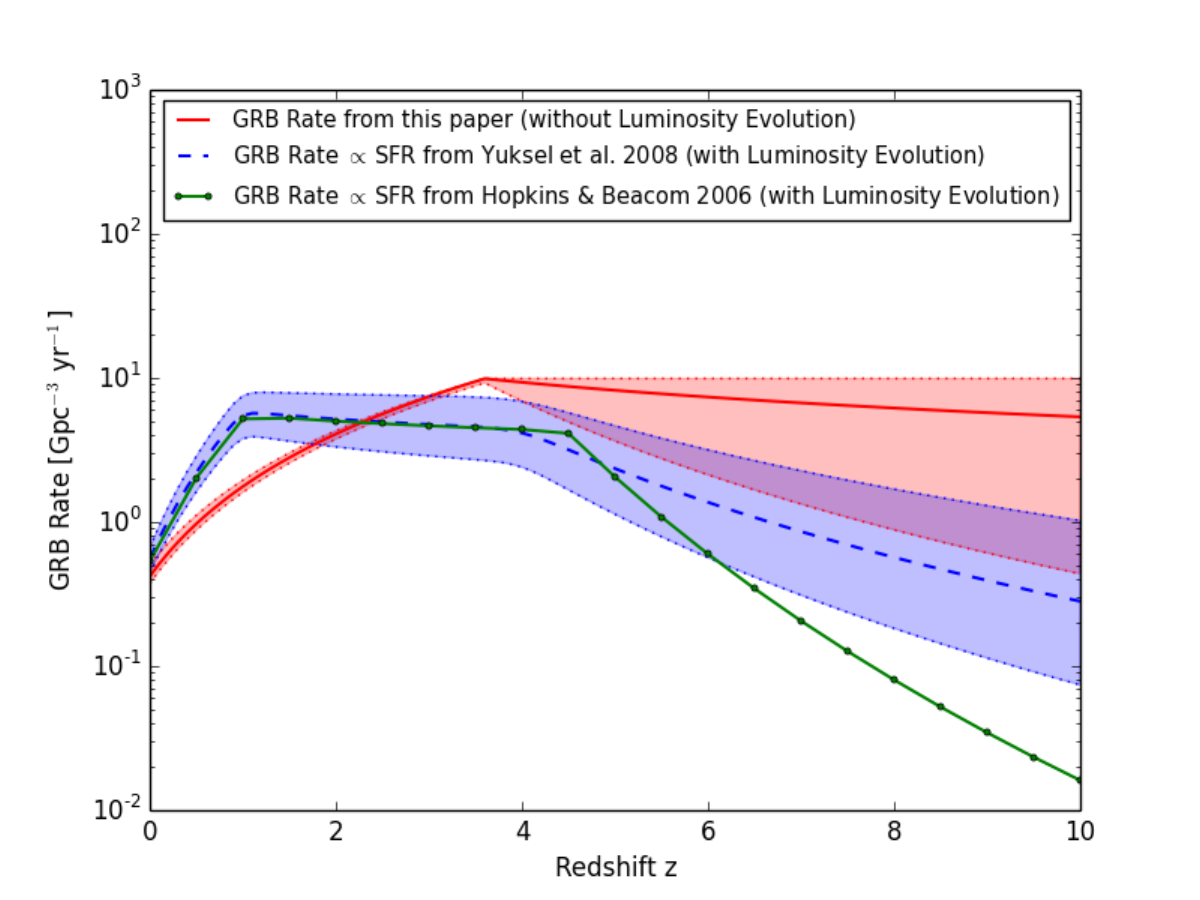}
\end{center}
\caption{
Comparison between the cosmic GRB rate from our best-fit sample (red line; Sect.~\ref{sect:best_result})
and the cosmic GRB rates that follow strictly the shapes of the SFRs from
\citet{Hopkins06} (green line) and \citet{Yuksel08} (blue line).
The GRB rates that trace the SFRs can generate results that match well with observational data
if luminosity evolution is included (see discussion in Sect.~\ref{sect:lum_evo}).
}
\label{fig:snr-grb}
\end{figure}

The comparison between our best-fit cosmic GRB rate and the shape of cosmic SFR
from both \citet{Hopkins06} and \citet{Yuksel08} can be found in Fig.~\ref{fig:snr-grb}.
Our best-fit GRB rate is plotted as a red line.
The green and blue lines in the figure show the GRB rates that strictly follow
the shapes of the SFR from \citet{Hopkins06} and \citet{Yuksel08}, respectively.
The normalizations of the green and blue lines come from 
fitting with real GRB observations by including luminosity evolution 
(see discussions in the following section, Sect.~\ref{sect:lum_evo}).

The red shaded region in Fig.~\ref{fig:snr-grb} shows the uncertainty of 
our best-fit GRB rate. We quantify the uncertainty by modifying the parameters 
in the GRB rate function (Eq.~\ref{eq:GRBrate}) around our best-fit set of parameters (Table~\ref{tab:best_result})
until results no longer match well with observations.
The shaded region in the figure indicates the parameter space
that produces results which satisfy the following three criteria:
(1) Matching with the observed peak-flux distribution with KS test significance level $> 90\%$.
(2) Matching with the observed redshift distribution with KS test significance level $> 90\%$.
(3) Prediction for the {\it Swift} detection rate within the range of $95 \pm \sqrt{95} = 95 \pm 10$ events per year.
Table~\ref{tab:best_result_lowlim} and \ref{tab:best_result_uplim} summarize the parameters and results 
for the lower and upper limit of the GRB rate shown in Fig.~\ref{fig:snr-grb}.
Note that for the peak-flux distribution match, we require the KS test significance including the error bars exceeds $90\%$.
Therefore, it is the upper limit of the KS test significance that needs to exceed $90\%$, 
as seen in the tables.

Additionally, the constraining factor for this uncertainty region of the GRB rate turns out 
to be the prediction of the Swift detections (i.e., the third criterion listed above), rather than the shapes of the functions.
In other words, all the GRB rates within the red-shaded region produce decent matches with the 
shapes of the observed peak-flux and redshift distributions. However, if we move the curve 
further away from the lower/upper limit of the red-shaded region, 
we will have the predicted detection rates lower/higher than the true rate of {\it Swift},
even though the shapes of the peak-flux and redshift distributions still match well with 
the observations.

\begin{table}[h]
\caption{\label{tab:best_result_lowlim}
Summary of the set of parameters for the lower limit of the GRB rate. 
Parameters for the luminosity function are the same as those shown in Table~\ref{tab:best_result}.
}
\begin{center}
\begin{tabular}{|c|c|c|c|c|c|}
\hline\hline
$R_{\rm GRB}(z=0)$        & $z_1$ & $n_1$ & $n_2$ & Detection & Prediction for \\
$ [\rm Gpc^{-3} \ yr^{-1}]$ &             &              &             & Rate         & {\it Swift} [$\rm yr^{-1}$] \\
\hline
0.38 & 3.60 & 2.10 & -3.50 & 18.70\% & 83.85 \\
\hline\hline
\end{tabular}
\end{center}

\begin{center}
\begin{tabular}{|c|c|c|c|}
\hline\hline
KS-test (D) for & Significance for & KS-test (D) for & Significance for \\
z distribution & z distribution & peak flux distribution & peak flux distribution \\
\hline
$5.70 \times 10^{-2}$ & 98.68\% & $3.32^{+2.56}_{-1.01} \times 10^{-2}$ & $93.09^{+6.80}_{-61.41}\%$ \\
\hline\hline
\end{tabular}
\end{center}

\end{table}

\begin{table}[h]
\caption{\label{tab:best_result_uplim}
Summary of the set of parameters for the upper limit of the GRB rate.
Parameters for the luminosity function are the same as those shown in Table~\ref{tab:best_result}.
}
\begin{center}
\begin{tabular}{|c|c|c|c|c|c|}
\hline\hline
$R_{\rm GRB}(z=0)$        & $z_1$ & $n_1$ & $n_2$ & Detection & Prediction for                  \\
$ [\rm Gpc^{-3} \ yr^{-1}]$ &             &              &             & Rate         & {\it Swift} [$\rm yr^{-1}$]  \\
\hline
0.51 & 3.60 & 1.95 & -0.00 & 12.95\% & 104.74  \\
\hline\hline
\end{tabular}
\end{center}

\begin{center}
\begin{tabular}{|c|c|c|c|}
\hline\hline
KS-test (D) for & Significance for & KS-test (D) for & Significance for \\
z distribution & z distribution & peak flux distribution & peak flux distribution \\
\hline
$5.41 \times 10^{-2}$ & 99.43\% & $5.12^{+3.98}_{-1.49} \times 10^{-2}$ & $58.10^{+34.07}_{-53.72}\%$ \\
\hline\hline
\end{tabular}
\end{center}

\end{table}

\citet{Horiuchi09} quantify the uncertainty of the star-formation rate measurements summarized in \citet{Yuksel08} 
by taking into account the scatter of data. In order to compare with our best-fit sample
and have a better idea of how well we can use the GRB rate to constraint the star-formation rate at high redshift, 
we plot this uncertainty as the blue shaded region in Fig.~\ref{fig:snr-grb},
with the same normalization factor used for the blue line. 
Note that the blue-shaded region is the uncertainty for the star-formation rate measurements, instead of the GRB rate. 
In other words, although the blue line can generate a good fit with the GRB observations using the current normalization 
if including luminosity evolution (see Sect.~\ref{sect:lum_evo}), there is no guarantee that GRB rates within the blue-shaded region
can all generate good matches with observations. 

Results in Fig.~\ref{fig:snr-grb} show a clear diversity in the shapes of the GRB rate versus redshift from the SFRs at 
$z \sim 4$. Both the GRB rate and the SFRs start decreasing beyond $z \sim 4$.
However, the SFRs, even the one from \cite{Yuksel08}, decline                                                                                                                                                                                                                                                                                                                                                                                                              
much faster than the GRB rate found in our best-fit sample. 
The SFRs from \citet{Hopkins06} and \citet{Yuksel08} decrease at large redshift with a power-law index of $\sim -8.0$
and $\sim -3.5$, respectively,
while the GRB rate found in this work decreases with a power-law index $\sim -0.7$.

There are several possible explanations for this very high GRB rate at large redshift.
First, this could suggest an even larger SFR at high redshift, 
which implies that most of the star formation activities at high redshift probably come from low-luminosity galaxies,
and thus measurements of the SFR based on galaxy observations might underestimate the 
true rate \citep[e.g.,][]{Kistler09, Jakobsson12, Tanvir12, Kistler13, Trenti13}.
Alternatively, a high GRB rate at early times can also be explained if the fraction of GRB-related supernovae
changes as a function of redshift. That is, if there are more supernovae accompanied by GRBs at high redshift, 
one could get a higher GRB rate without adjusting the SFR.
For example, \citet{Woosley12} suggest several collapsar models that can
generate long gamma-ray transients, and state that these events are more frequent 
in the early universe.

Another possibility would be the luminosity evolution. 
If we allow luminosity evolution in our simulation
and generate more high-luminosity bursts at high redshift than at low redshift, this could 
create enough low-flux bursts without over-producing the detections at low redshift.
In this case, we might not need a high GRB rate at large redshift to balance the low-flux ratio 
of those from the observed GRBs. 
Several studies have already suggested the possibility of redshift evolution of the GRB luminosity function
\citep[e.g.,][]{Salvaterra09, Salvaterra12, Virgili11, Toma11,Kanaan13}. We will thus investigate this possibility in the following section.

\section{Possibility of Luminosity Evolution}
\label{sect:lum_evo}

To explore whether the extreme excess of GRB rate at high redshift is not necessary when 
luminosity evolution is considered,
we test several functions for luminosity evolution using the shapes of the previously reported SFRs 
as the intrinsic GRB rate.
Our goal is to study whether it is possible to generate a sample that matches well with
real observations, while having the GRB rate restricted to follow the current SFR
measurements.
Specifically, we perform the tests with two possible SFRs: 
(1) the commonly adopted SFR from \cite{Hopkins06}, 
and (2) the one in \citet{Yuksel08}, which is similar to the first one but with high-redshift corrections using GRB detections.

For simplicity, we made the characteristic luminosity $L_{\star}$ in
the luminosity function (Eq.~\ref{eq:GRBlum}) change as a function of redshift $z$.
Two different functional forms are tested in our simulations: (1) $L_{\star} = A \times z^{B}$, and 
(2) $L_{\star} = A \times \rm log_{10}(z)$.
Again, all simulations with luminosity evolution are based on 26884 active detectors,
which is the average number from year 2005 to year 2009.

\begin{table}[!h]
\caption{\label{tab:lum_evo_Hopkins}
Summary of the parameters and results of the best-fit sample with luminosity evolution using the SFR from \citet{Hopkins06}.
}
\begin{center}
\begin{tabular}{|c|c|}
\hline\hline
$R_{\rm GRB}(z=0) [\rm Gpc^{-3} \ yr^{-1}]$ & Functional Form \\
\hline
0.54 & SFR in \citet{Hopkins06} \\
\hline\hline
\end{tabular}
\end{center}

\begin{center}
\begin{tabular}{|c|c|c|c|c|c|c|}
\hline\hline
$L_\star$                   & A & x & y & $E^{\rm src}_{\rm peak}$ & Detection & Prediction for \\
$[\rm erg \ s^{-1}]$   &     &    &    &   distribution                        & Rate         & {\it Swift} [$\rm yr^{-1}$] \\
\hline
$10^{51.00}$ & 2.0 & -0.20 & -2.00 & Flat in Log Space & 22.28\%& 95.27 \\
\hline\hline
\end{tabular}
\end{center}

\begin{center}
\begin{tabular}{|c|c|c|c|}
\hline\hline
KS-test (D) for & Significance for & KS-test (D) for & Significance for \\
z distribution & z distribution & peak flux distribution & peak flux distribution \\
\hline
$5.96 \times 10^{-2}$ & 97.26\% & $2.42^{+3.21}_{-0.59} \times 10^{-2}$ & $98.76^{+1.22}_{-76.28}\%$ \\
\hline\hline
\end{tabular}
\end{center}

\end{table}

\begin{table}[!h]
\caption{\label{tab:lum_evo_Yuksel}
Summary of the parameters and results of the best-fit sample with luminosity evolution using the SFR from \citet{Yuksel08}.
}
\begin{center}
\begin{tabular}{|c|c|}
\hline\hline
$R_{\rm GRB}(z=0) [\rm Gpc^{-3} \ yr^{-1}]$ & Functional Form \\
\hline
0.56 & SFR in \citet{Yuksel08} \\
\hline\hline
\end{tabular}
\end{center}

\begin{center}
\begin{tabular}{|c|c|c|c|c|c|c|}
\hline\hline
$L_\star$                   & A & x & y & $E^{\rm src}_{\rm peak}$ & Detection & Prediction for \\
$[\rm erg \ s^{-1}]$   &     &    &    &   distribution                        & Rate         & {\it Swift} [$\rm yr^{-1}$] \\
\hline
$10^{51.00}$ & 1.9 & -0.20 & -2.00 & Flat in Log Space & 21.01\%& 95.11 \\
\hline\hline
\end{tabular}
\end{center}

\begin{center}
\begin{tabular}{|c|c|c|c|}
\hline\hline
KS-test (D) for & Significance for & KS-test (D) for & Significance for \\
z distribution & z distribution & peak flux distribution & peak flux distribution \\
\hline
$6.13 \times 10^{-2}$ & 96.48\% & $2.34^{+3.76}_{-0.61} \times 10^{-2}$ & $99.19^{+0.80}_{-83.49}\%$ \\
\hline\hline
\end{tabular}
\end{center}

\end{table}

Results show that the second form ($L_{\star} \propto \rm log_{10}(z)$) can create mock-triggered samples
that match well with the observations,
with some adjustments of the parameters in the luminosity function.
Table~\ref{tab:lum_evo_Hopkins} and \ref{tab:lum_evo_Yuksel}
summarize
the GRB characteristics that generate a good match with observations
when adopting luminosity evolution and the SFR from \citet{Hopkins06} and \citet{Yuksel08}, respectively.
The KS-test values for these two samples comparing with the observed redshift and peak-flux distributions
are also given in the tables.
As expected, a more severe luminosity evolution is needed if assuming the SFR from \citet{Hopkins06},
because this rate decreases more rapidly at high redshift than that from \citet{Yuksel08}.

The green and blue lines in Fig.~\ref{fig:snr-grb}
plot these two GRB rates with characteristics summarized in Table~\ref{tab:lum_evo_Hopkins} and \ref{tab:lum_evo_Yuksel}, respectively. 
As discussed before, the GRB rate shown as the green line strictly follows the shape of the SFR from \citet{Hopkins06},
while the GRB rate in blue traces the shape of the SFR from \citet{Yuksel08}.
Our search suggests that these two GRB rates can only match well with real GRB observations if
luminosity evolution is included.

\section{Discussions and Conclusions}
\label{sect:conclusions}

We developed a program that simulates the complex trigger algorithm adopted by BAT.
We used this program to search for a cosmic GRB rate and luminosity function
that generates a mock-triggered sample with characteristics that match well with observations. 
Our results suggest that the BAT's complex trigger algorithm 
increases the detection rate by $\sim 1.2$ to 5.3 times higher than that using a single flux threshold. 
Therefore, adopting the complex trigger algorithm of BAT improves the chance of triggering bursts with low fluxes.
As a result, we need an intrinsic GRB sample that is on average dimmer
than previously expected, in order to avoid over-producing the number of detections and to match with real {\it Swift} observations. 
This can be achieved by either adding more bursts with lower luminosities,
increasing the number of bursts at high redshift, or both.

According to all the parameter sets we tried, generating more bursts with lower luminosities 
in the intrinsic GRB luminosity function
has a side effect of
creating too many detections at low redshift, and thus resulting in a distribution that does not
match well with the redshift distribution of the real BAT-detected GRBs.
Therefore, adding more bursts at large redshift provides a way to
increase the number of dim bursts without over-producing low-redshift observations,
and thus generates a mock-triggered sample that matches well with
both the redshift and the peak-flux distributions of real observations.

By adopting the complex trigger algorithm of BAT, 
the best-fit sample we found suggests the possibility of an intrinsic GRB rate that contains even more bursts at higher redshift 
than previous expectations \citep{Yuksel08, Butler10, Wanderman10}.
This result implies that either (1) the SFR in the early universe is even higher than
that inferred by previous GRB studies, and hence
a majority of star formation might happen in low-luminosity galaxies,
or (2) some redshift evolution effects, such as luminosity evolution or an evolution in the GRB-to-supernovae ratio, need to be taken into account. 

Theoretical studies suggest that the GRB characteristics are likely to be different in the early universe 
\citep[e.g.,][]{Salvaterra09, Virgili11, Toma11, Woosley12}.
Therefore, at least some redshift evolution in the GRB luminosity functions are expected.
We thus examine the possibility of including luminosity evolution to 
reduce the intrinsic GRB rate at high redshift 
and maintain a good match with observations.
We found that if we assume the shape of the SFR from \citet{Yuksel08},
the characteristic luminosity $L_{\star}$ in Eq.~\ref{eq:GRBlum}
needs to evolve as $L_{\star} = 1.9 \times \rm log_{10}(z)$ in order to generate 
a mock-triggered sample that matches well with both 
the observed redshift and peak-flux distributions.
If we assume the shape of the SFR from \citet{Hopkins06},
which predicts an even lower GRB rate at high redshift,
a more severe luminosity evolution with $L_{\star} = 2.0 \times \rm log_{10}(z)$
is needed to produce a good match with observations.

Besides the GRB luminosity,
other GRB characteristics might also evolve with redshift.
Theoretical studies suggest that low-metallicity population III stars produce GRBs
that have much longer durations than regular ones \citep{Fryer01, Komissarov10, Meszaros10, Toma11, Woosley12}. 
For example, \citet{Gendre13} propose that the ultra-long GRB 111209A,
which had a prompt emission that lasted around $1.5 \times 10^4$ seconds, 
resulted from a low-metallicity blue supergiant star and 
resembled more the population III star explosions.
Therefore, one might expect GRBs in the early universe to have longer durations.
The BAT trigger algorithm, particularly the rate trigger, is relatively insensitive to these ultra-long bursts 
due to their slow-changing light curve.
Our trigger simulator shows that an ultra-long burst like GRB 111209A
(with similar pulse duration, $E_{\rm iso}$, and spectral parameters)
can be detected by image trigger out to redshift $z \sim 1.17$ if
the event happens on-axis relative to the detector plane (Grid ID = 17),
and $z \sim 0.3$ if the burst is far off-axis (Grid ID = 14).
The rate trigger criteria can only detect such bursts within 
$z \sim 0.17$ even if the burst appears on axis.
Note that in this particular simulation, we still assume a flat background level
throughout the burst duration, which is likely not to be true when the 
event lasts for a few hours. 
Therefore, such long-duration GRBs are even harder to be detected  
by the BAT in practice, and thus might introduce further uncertainty of 
the intrinsic GRB rate at high redshift.

The best-fit sample from our simulations suggests the GRB rate in the local universe to be $\sim 0.42 \ \rm Gpc^{-3} \ yr^{-1}$,
which is in general agreement (within a factor of 2) with other observations and studies 
\citep[e.g.,][]{Schmidt01,Guetta07_Ibc,Liang07,Pelangeon08,Wanderman10}.
Note that this is the rate of GRBs beamed toward us, as we do not consider bursts that are pointed away from us,
due to the large uncertainty in the beaming factor
(see discussion in Sect~\ref{sect:intro}).
In addition, the best-fit sample suggests that BAT should detect $\sim 96$ GRBs per year,
which is consistent with the real BAT detection of $\sim 95$ bursts per year averaged from 2005 to 2009 \citep{Sakamoto11}.
According to our best-fit sample, $\sim 1\%$ of all detections (i.e., $\sim 1$ burst per year) come from redshift $z \gtrsim 6$.

Moreover, the GRB rate from our best result gives a total number of $4571^{+829}_{-1584}$ GRBs per year that are beamed toward us
in the whole universe.
The errors here are calculated using the lower and upper limits of the GRB rate 
shown in Fig.~\ref{fig:snr-grb} (the red-shaded region).
To have a better idea how this number is compares to the cosmic star-formation history
and the core-collapse supernova rate,
we perform a simple order-of-magnitude calculation, as described below.
Using the current star-formation rate measurements \citep[e.g.,][]{Hopkins06,Yuksel08} 
and the commonly-adopted modified Salpeter Initial Mass Function \citep{salpeter,bg},
one can estimate
the total core-collapse supernova rate in the whole universe to be $\sim 10^8$ per year \citep[for example, see calculation in][]{lf}.
Observations suggest that $\sim 25\%$ of all the core-collapse supernovae are Type Ibc event \citep{Li11b},
and only $\sim 1\%$ of all Type Ibc supernovae are related to GRBs \citep[e.g.,][]{Berger03}.
An order-of-magnitude calculation then gives $\sim 2.5 \times 10^5$ GRB per year in the whole universe.
If we assume the beaming factor to be $\sim 50$ \citep[i.e., 1/50 GRBs are pointed at us;][]{Guetta05}, 
there will be $\sim 5000$ GRBs per year that are beamed toward us,
which is consistent with our result.

Furthermore, for all the different $E^{\rm src}_{\rm peak}$ distributions we tried,
it seems to be difficult to generate results that match well with the observed $E^{\rm obs}_{\rm peak}$
distribution without assuming some kind of intrinsic correlation between $E^{\rm src}_{\rm peak}$
and the burst energy output, such as the $L_{\rm peak}$. Similar conclusions have been drawn
by many groups that study possible correlations in GRB characteristics \citep[e.g.,][]{Ghirlanda12,Shahmoradi13}. 
Better $E^{\rm obs}_{\rm peak}$ measurements will be crucial 
to verify this conclusion.
 
We presented here a GRB rate and luminosity function that generates a mock-trigger sample that matches well with observation.
To confirm whether there exist equally good or better fits
other than the one we presented, and also to quantify the probability of whether 
the good match is due to a real physical solution or just a coincidental match,
a complete Monte Carlo search would be required. 
Therefore, a significant decrease of the simulation time will be important 
to further pin down the cosmic GRB rate, improve our understanding of the GRB characteristics,
and quantify their uncertainties.

Observationally, increasing the number of GRB detections with redshift measurement 
is essential in reducing the uncertainties in the observed distributions 
and better constraining the GRB properties.
Moreover, better knowledge of the biases in GRB follow-up observations
and redshift determination is critical to understand the completeness of 
the observational GRB sample.  
The rapidly growing number of GRB detections by {\it Swift}
and the effort in measuring GRB characteristics through multi-wavelength observations
will certainly improve our understanding of GRB physics, and will make them even better probes of stellar evolution
and star-formation history out to the early universe.

\noindent{\bf{Acknowledgments}}\\
We are grateful for valuable discussions with Brian Fields,
Brett Hayes, Daniel Kocevski, Judith Racusin, Jon Hakkila, Amir Shahmoradi,
Lorenzo Amati, John Beacom, Jay Cummings, Antonino Cucchiara, and Dieter Hartmann.
We also appreciate the helpful comments and suggestions from the anonymous referee.
Amy Lien is supported by an appointment to the NASA Postdoctoral Program at 
the Goddard Space Flight Center, administered by Oak 
Ridge Associated Universities through a contract with NASA.
We thank Sarah Antier for pointing out a typo in Eq.~\ref{eq:Yonetoku_mod}.

\appendix
\section{Erratum: Probing the Cosmic Gamma-Ray Burst Rate with Trigger Simulations of the \textit{Swift} Burst Alert Telescope}

This is an erratum to the paper ``Probing the Cosmic Gamma-Ray Burst Rate with Trigger Simulations of the \textit{Swift} Burst Alert Telescope'' that was published in ApJ, 783, 24L (2014). Recently we found a mistake in the code, which affects the normalization of the GRB rate, i.e., the parameter $R_{\rm GRB}(z=0)$ in the following equation (Eq. 1 in the original paper): 

\beq
\label{eq:GRBrate}
R_{\rm GRB}(z) = R_{\rm GRB}(z=0) \ \left \{  \begin{array}{ll}
(1+z)^{n_1}, & z \leq z_1,\\
(1+z_1)^{n_1-n_2} \ (1+z)^{n_2}, & z>z_1 \end{array} \right.
\eeq 

The mistake was caused by incorrectly using the numerical recipe subroutine \citep{NR92}, ``qromb'', recursively to perform a double integration. Because one of the routines (``trapzd'') called by qromb uses a static local variable, this function should not be used recursively. 

For the equation we use for GRB rate calculation, incorrectly using ``qromb'' recursively results in an integrated value that is a factor of two smaller than the correct number. In other words, the true integrated number is twice as high as what we calculated. Therefore, we need to lower the normalization parameter $R_{\rm GRB}(z=0)$ by a factor of two in order to have the same accumulated GRB number that matches with $\it Swift$'s detection. This affects all the numbers of $R_{\rm GRB}(z=0)$ reported in the paper (all of them need to be divided by 2), including the normalization parameter in our best-fit model (mentioned in Table 2 and Section 11 in the paper), in the the upper and lower limit (Table 5 and 6 in the paper), and in the fits we used to study the luminosity evolution (Table 7 and 8 in the paper). The corrected $R_{\rm GRB}(z=0)$ are listed in Table~\ref{tab:corrected_result} below. In addition, the corrected version of Fig. 17 with all the normalizations lower by a factor of 2 is also updated in this paper version. The Appendix in this erratum (Appendix B) includes a more detailed explanation of where the double integration and the factor of two come from.

\begin{table}[h]
\caption{\label{tab:corrected_result}
Summary of the normalization parameters $R_{\rm GRB}(z=0)$ presented in the paper that are affected by this mistake. The corrected 
numbers here are half of the original numbers.
}
\begin{center}
\begin{tabular}{|c|c|}
\hline\hline
Location in the Original Paper & Corrected $R_{\rm GRB}(z=0) [\rm Gpc^{-3} \ yr^{-1}]$ \\
\hline
Table 2 and Section 11 & 0.42 \\
Table 5 & 0.38 \\
Table 6 & 0.51 \\
Table 7 & 0.54 \\
Table 8 & 0.56 \\
\hline\hline
\end{tabular}
\end{center}

\end{table}

To double check for consistency, we use Python and the Scipy library \citep[the ``quad'' subroutine;][]{SciPy_ref} to perform the integration. We found the integrated GRB number (i.e., the total number of GRBs in the Universe that are beamed toward us) to be $4571_{-1584}^{+829}$ GRBs per year. This number is very similar to the one reported in the original paper, as expected. The small difference is likely due to numerical uncertainty and rounding error.
  
We have double checked all of our codes to make sure the normalization is the only thing affected by this problem. We sincerely apologize for this mistake, and for not noticing it earlier. 

\noindent{\bf{Acknowledgments}}\\
We thank Philip Graff for pointing out the mistake. Amy Lien is grateful for the help from Philip Graff, Brett Hayes, David Friedlander, Craig Markwardt, James Bubeck, Mike Arida, and all the collaborators on this paper to further pinpoint the problem.



\section{Erratum Appendix: Detailed explanation of the problem}

We intergrate Eq. 2 in the paper
\beq
\label{eq:GRBrate_dz}
R_{\rm GRB;dz}(z) = \frac{dN}{d\Omega \ dz \ dt_{\rm obs}} = R_{\rm GRB}(z) \ \frac{dV_{\rm comov} dt_{\rm src}}{d\Omega dz dt_{\rm obs}} = \frac{R_{\rm GRB}(z)}{(1+z)} \ \frac{dV_{\rm comov}}{d\Omega dz}
\eeq
from redshift $z=0$ to $z=10$ to get the total GRB number in the Universe (per solid angle per observation time), and convert this intrinsic rate to a detected rate by multiplying the detection fraction, the {\it Swift}/BAT's field of view, and the {\it Swift} survey time (see Eq. 9 in the paper). The estimated detection rate is then used for constraining the normalization factor $R_{\rm GRB}(z=0)$ by comparison with the true {\it Swift} detection rate. Equation 2 has an implicit integration from $dV_{\rm comov} = r^2_{\rm com} \ d{\Omega} \ dr_{\rm com}$, where $r_{\rm com} = \int (c/H(z)) dz$. $H(z)$ is the Hubble parameter and c is the speed of light.  

\begin{table}[!h]
\caption{\label{tab:corrected_result_test}
An example of how the integrated numbers are affected by this mistake, with different integrated redshift ranges. The ratio in the fourth column refers to the ratio
of the incorrect number over the correct number. This example uses the original incorrect $R_{\rm GRB}(z=0)$, and the numbers here are only for demonstration purpose.
}
\begin{center}
\begin{tabular}{|c|c|c|c|}
\hline\hline
Integrated redshift range & Correct number & Incorrect number & Ratio \\
\hline
0.0-0.5 & 32.45 & 19.02 & 0.586 \\
0.0-1.0 & 225.12 & 117.48 & 0.522 \\
0.0-1.5 & 631.18 & 322.12 & 0.510 \\
0.0-2.0 & 1233.58 & 624.56 & 0.506 \\
0.0-2.5 & 1997.64 & 1007.58 & 0.504 \\
0.0-3.0 & 2890.59 & 1454.86 & 0.503 \\
0.0-3.5 & 3885.70 & 1953.10 & 0.503 \\
0.0-4.0 & 4867.24 & 2444.45 & 0.502 \\
0.0-4.5 & 5660.97 & 2841.78 & 0.502 \\
0.0-5.0 & 6306.03 & 3164.78 & 0.502 \\
0.0-5.5 & 6836.13 & 3430.21 & 0.502 \\
0.0-6.0 & 7276.30 & 3650.65 & 0.502 \\
0.0-6.5 & 7645.34 & 3835.48 & 0.502 \\
0.0-7.0 & 7957.56 & 3991.86 & 0.502 \\
0.0-7.5 & 8223.80 & 4125.13 & 0.502 \\
0.0-8.0 & 8452.59 & 4239.77 & 0.502 \\
0.0-8.5 & 8650.59 & 4339.09 & 0.502 \\
0.0-9.0 & 8822.93 & 4425.49 & 0.502 \\
0.0-9.5 & 8973.78 & 4500.84 & 0.502 \\
0.0-10.0 & 9106.65 & 4567.65 & 0.502 \\
\hline\hline
\end{tabular}
\end{center}

\end{table}

Because of how the numerical recipe subroutine ``trapzd'' is set up, as long as the number of the first integration ($\int R_{\rm GRB;dz}(z) dz$ in our case) is much greater than the second integration ($\int (1/H(z)) dz$ in our case), incorrectly using the subroutine recursively will always produce a number about a factor of 2 smaller than the correct answer. Table~\ref{tab:corrected_result_test} shows an example of the fraction of incorrect and correct answer converges to $\sim 0.5$ as we integrate with a larger redshift range. The numbers in this example are calculated from the GRB parameters of Table 2 in the paper, including the original incorrect $R_{\rm GRB}(z=0)$.

\bibliographystyle{apj}
\bibliography{ref}

\end{document}